\documentclass[journal,11pt,draftclsnofoot,onecolumn]{IEEEtran}

\usepackage{setspace}
\usepackage{cite}
\usepackage{eucal}
\usepackage{amsfonts}
\usepackage{pifont}
\usepackage[pdftex]{graphicx}
\usepackage[cmex10]{amsmath}
\usepackage{ dsfont }%
\usepackage[ruled, vlined]{algorithm2e}
\usepackage[caption=false,font=footnotesize]{subfig}
\usepackage{fixltx2e}
\usepackage{pbox}

\usepackage{url}

\allowdisplaybreaks

\begin{document}

\title{Distributed Protocols for Interference Management in Cooperative Networks}

\author{Christopher~Hunter,~\IEEEmembership{Student Member,~IEEE,}
        Ashutosh~Sabharwal,~\IEEEmembership{Senior Member,~IEEE}
        
\thanks{C. Hunter and A. Sabharwal are with the Department
of Electrical and Computer Engineering, Rice University, Houston,
TX, 77005 USA e-mail: [chunter,ashu]@rice.edu. This work was partially funded by NSF grants CNS-0551692, CNS-0619767, CNS-0923479
and CNS-1012921.}}%

\maketitle

\begin{abstract}
In scenarios where devices are too small to support MIMO antenna arrays, symbol-level cooperation may be used to pool the resources of distributed single-antenna devices to create a virtual MIMO antenna array. We address design fundamentals for distributed cooperative protocols where relays have an incomplete view of network information. A key issue in distributed networks is potential loss in spatial reuse due to the increased radio footprint of flows with cooperative relays. Hence, local gains from cooperation have to balance against network level losses. By using a novel binary network model that simplifies the space over which cooperative protocols must be designed, we develop a mechanism for the systematic and computational development of cooperative protocols as functions of the amount of network state information available at relay nodes. Through extensive network analysis and simulations, we demonstrate the successful application of this method to a series of protocols that span a range of network information availability at cooperative relays.
\end{abstract}

\section{Introduction}
\begin{figure}[h!]
	\centering
	\includegraphics[width=2.0in]{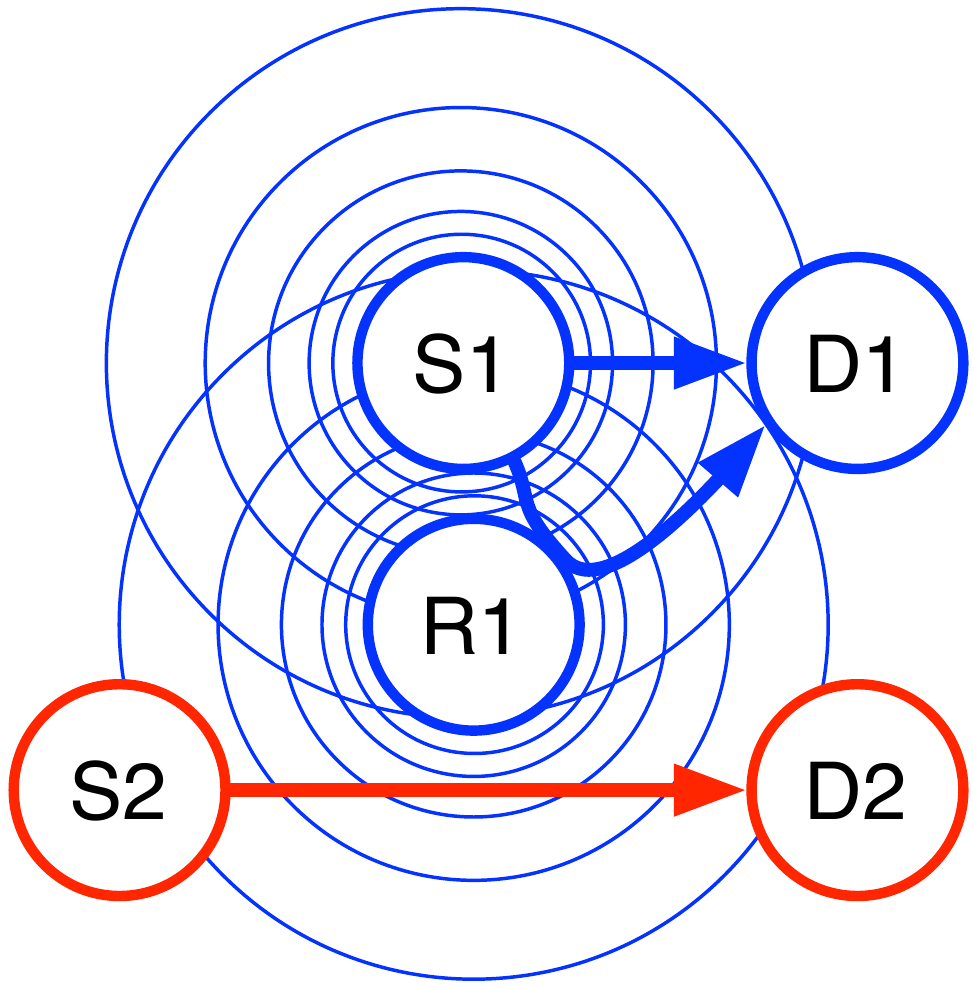}
	\caption{Relays can decrease spatial reuse by adding interference.}
	\label{fig:tradeoff}
\end{figure}

Symbol-level cooperation between neighboring wireless nodes is known to have the potential for large data-rate gains in wireless fading channels~\cite[and references therein]{laneman_coopdiv,kramer2006cc,liu2009cooperative,dohler2010cooperative}. However, cooperative transmissions also lead to increased radio footprints due to multiple simultaneous transmissions by nodes which are spatially distributed (see Figure~\ref{fig:tradeoff} as an example). Thus, it is possible that the increase in the throughput of one flow comes at the expense of reduced spatial reuse. The actual tradeoff depends on the form of inter-flow coordination protocols in the network, which determines the timing and form of nodes' transmissions. In turn, the form of coordination depends on how much network state information is available at each node. In this work, we systematically analyze the role of network state information and the associated design of random access protocols in the context of cooperative communication-based networks. 

Our core contribution is a technique for protocol development for cooperative relays. This technique applies for {\it any} subset of full network state information, which constitutes both the channel states of all links and node states. This contribution is framed by three key results. First, we propose a binary approximation for the network, simplifying all random variables to two-state variables. The binary approximation is then used to create relay access policies where the relay has zero, one, or two hops of channel information about the rest of the network. For each case of channel state information (zero, one, and two hops), we also consider the impact of whether the relay adopts a conservative or greedy viewpoint about the \emph{unknown} network state information. These access policies serve as guidelines to design cooperative protocols for actual wireless channels.

Second, we compare the partial information policies to the policy which has full information to quantify the performance impact of each piece of network state information. The six protocols (\{greedy or conservative\} relay $\times$ \{zero, one, or two\} hops of channel knowledge) quantify an intuitive result. If the relay is greedy and assumes the best case scenario about what is not known about the rest of the network, then gains can be significant for the cooperative flow but they come at the expense of significant loss for other flows in some topologies. In contrast, a conservative relay, which aims to cause no harm to other flows, requires a substantial amount of network information to provide any reasonable cooperative gains. In short, a relay can be both helpful and socially responsible only if it has significant information about the state of the network. Otherwise, a conservative relay will end up staying silent in the bulk of unknown network states in order to avoid any harm to a neighboring flow. 

Lastly, we close the loop by translating the relay access policies from the binary approximation to $\mathsf{SINR}$-based protocols and study their performance using hardware-accurate network simulations. This last step is possible for all but the single-hop knowledge policies because the binary collision model turns out to be too crude to predict the behavior in this case. For the other four protocols, our simulations reveal that the trends predicted by the binary model hold for the fading channels.

The proposed ``computational" mechanism for protocol design is inspired by the fact that designing medium access protocols with provable performance is often analytically and/or computationally intractable due to large state space. As a result, medium access protocols are often designed on a case-by-case basis with different amounts of network state information. As a notable exception, the authors in~\cite[and references therein]{chiang2007layering} reverse engineer the exponential backoff structure of many random-access MAC protocols as a solution to a non-cooperative game. Similarly, the authors in~\cite{rodoplu2008challenges} present an optimization-based framework for automated protocol design that solves an example scheduling problem. These works pursue a different methodology to a similar high-level goal: the construction of protocols in a procedural fashion.

Our methodology is similar in spirit to recent work on deterministic approximation information theoretic analyses~\cite{avestimehr2011wireless}, where deterministic network models provide an insight into the design of Gaussian network models in many, but not all, cases---for example, the deterministic model in~\cite{avestimehr2011wireless} is not a useful approximation of the MIMO channel. 

Information theoretic analyses of cooperative communication have a sizable body of literature~\cite[and references therein]{laneman_coopdiv,kramer2006cc,liu2009interference}. These works generally assume perfect network knowledge and centralized coordination to establish performance bounds on cooperative networks. In practice, global network knowledge at every node is often infeasible and/or not scalable as the size of the network increases. In contrast, we study distributed cooperative protocols. The work in~\cite{lichte2009analyzing} studies the effects of interference and cooperative networks from the opposite perspective of our work. Whereas~\cite{lichte2009analyzing} studies the implications of interference on the performance of a cooperative flow, we design protocols that address the implications of increased interference by cooperative flows on the rest of the network.

Network-layer distributed cooperative protocols have a considerably more sparse presence in the literature. A survey of the current state-of-the-art~\cite{shan2009distributed} highlights the spatial reuse issue as an open problem. The protocols in~\cite{zhou2011novel, adam2010core,lu2009design ,liu2007ccm,zhu2005rre} rely on non-simultaneous transmissions, and hence are a form of opportunistic routing. We are interested in protocols where the source and relay transmit simultaneously within the same bandwidth since these simultaneous transmission schemes achieve higher rates~\cite{nabar2004performance} and simplify transceiver design~\cite{Hunter:2010,warpOFDM_assilomar} compared to their non-simultaneous counterparts. The protocols in~\cite{gokturk2008cooperation,jakllari} use RTS/CTS packet exchanges to mitigate interference caused by relaying on surrounding flows. RTS/CTS is disabled by default in the majority of 802.11 chipsets because of the overhead suffered on every transmission. In contrast, we study {\it reactive} NACK-based cooperative protocols that do not require preemptive handshakes to coordinate cooperation.

While not specifically targeting cooperation applications, there exists a sizable body of literature on managing interference in ad hoc networks. These varied strategies range from altering carrier-sense thresholds according to network dynamics~\cite{vasan2005echos}, to modifying the NAV structure of 802.11 to be less conservative~\cite{monks2001power,cesana2003interference}, and finally to using out-of-band busy tones to enhance channel reservations~\cite{haas2002dual}. To address the main challenge in managing relay-induced interference, we have chosen to base our protocol design on standard CSMA/CA access mechanisms like the IEEE 802.11 DCF. Conceptually, we believe that the prior literature can be leveraged in the context of cooperative interference management by using the proposed framework.

The rest of the paper is organized as follows: In Section~\ref{sct:model}, we describe our signal, decoding, and carrier-sense model. In Section~\ref{sct:policy}, we construct a binary approximation of this model and develop relaying policies for different amounts of network state information. We evaluate the relative performance of these relaying policies by considering their propensity to assist or harm the network. In Section~\ref{sct:protocols}, we translate these policies into $\mathsf{SINR}$-based protocols and evaluate their performance using a custom network simulator.

\section{System Model}
\label{sct:model}

In this section, we describe our signal model,  decoding model, and carrier-sensing model. We then discuss physical layer relaying schemes and define the desired relaying policies. 

\subsubsection{Signal Model}

We assume a slow fading model on the propagation of wireless signals. The reception of a transmission from a source node $\mathsf{S}$ at a destination node $\mathsf{D}$ in the presence of interferers is represented by
\begin{equation}
y_{\mathsf{D}} = h_{\mathsf{S}\mathsf{D}}x_{\mathsf{S}} + \sum_{i\in\mathcal{I}}h_{i\mathsf{D}}x_{i} + z_{\mathsf{D}},
\end{equation}
\noindent where $y_{\mathsf{D}}$ represents the received signal at $\mathsf{D}$ and $x_{\mathsf{S}}$ represents the transmitted signal from $\mathsf{S}$. The multiplicative fade $h_{ij}$ between nodes $i$ and $j$ remains constant for {\it at least} the duration of $x$.\footnote{In this formulation, we make no assumptions on the distribution from which $h_{ij}$ is drawn. In Section~\ref{sct:protocols} we will evaluate the protocols in a Rayleigh fading environment, but our proposed methodology applies to other channel distributions.} The additive noise $z_{\mathsf{D}}$ is assumed to be circularly symmetric complex Gaussian random variable that is drawn i.i.d. for every sample of $x_i$. The set $\mathcal{I}$ contains all other simultaneously transmitting sources in the network that act as interferers to $\mathsf{S}$.
\subsubsection{Decoding Model}
We further assume an $\mathsf{SINR}$-based decoding model that allows $\mathsf{D}$ to correctly decode a packet from $\mathsf{S}$ if and only if
\begin{equation}
\frac{H_{\mathsf{S}\mathsf{D}} E[|x_{\mathsf{S}}|^2]}{ \sum_{i\in\mathcal{I}}H_{i\mathsf{D}} E[|x_{i}|^2] + E[|z_{\mathsf{D}}|^2]} \geq \gamma_\mathsf{DEC},
\end{equation}
\noindent where $H_{ij}=|h_{ij}|^2= |h_{ji}|^2 $ represents the instantaneous, path-symmetric power of the fading channel, $E[\cdot]$ represents an expected value over the duration of the transmission $x_\mathsf{S}$, and $\gamma_\mathsf{DEC}$ is an $\mathsf{SINR}$ detection threshold.

\subsubsection{Carrier-sensing Model}
When a node $\mathsf{S}$ is backlogged with packets to send and is currently receiving, it will pause the state of its backoff counter when the total received energy exceeds a threshold, or
\begin{equation}
 \sum_{i\in\mathcal{I}}H_{i\mathsf{S}} E[|x_{i}|^2] + E[|z_{\mathsf{D}}|^2] \geq \gamma_\mathsf{CS},
\end{equation}
\noindent where $\gamma_\mathsf{CS}$ is a carrier-sensing power threshold.

\subsubsection{NACK-based Relaying Protocols}
Many schemes for cooperative signaling have been proposed. For example, the two most common methods for signaling that can improve diversity in reception over direct transmission are the Amplify-and-Forward (AF) and Decode-and-Forward (DF) schemes\footnote{In many of the works in the information theoretic literature (e.g. \cite{laneman_coopdiv}), these signaling schemes are referred to as protocols. In our work, we reserve the protocol terminology for higher-layer MAC behaviors and refer to these signaling methods as schemes.}~\cite{laneman_coopdiv}. In our prior work, we designed and implemented a NACK-based cooperative MAC layer~\cite{Hunter:2010} alongside a DF-capable cooperative PHY transceiver~\cite{murphy2011} in a real-time FPGA-based prototyping platform and showed large improvements in throughput and bit-error-rate. We use this implementation as the basis for the MAC layer protocol development in this paper. In principle, however, the discussion throughout this work can easily be extended to incorporate other signaling methods that are derived from AF and DF.

In a NACK-based cooperative MAC protocol, the relaying phase of cooperation is only engaged when a direct transmission between source and destination fails due to insufficient channel quality~\cite{shan2009distributed}. Synchronizing cooperative transmission to this event at both the source and relay simultaneously is solved by explicit NACK broadcast from the destination\cite{Hunter:2010}.
\subsubsection{Relaying Policies}
We refer to the instantaneous snapshot of network dynamics as network state information ($\mathsf{NSI}$). Consider a network of nodes represented by the set $\mathcal{N}$. Two key components frame $\mathsf{NSI}$,
\begin{equation}
\label{eq:continuousNSI}
\mathsf{NSI} := \left\{\mathcal{H}^{\frac{|\mathcal{N}|\left(|\mathcal{N}|-1\right)}{2}},\mathcal{X}^{|\mathcal{N}|} \right\},
\end{equation}
\noindent where
\begin{align}
\mathcal{H} &:=\{H_{ij} | \forall i,j \in \mathcal{N}, i\neq j\} \\
\mathcal{X} &:= \{X_i | \forall i\in \mathcal{N} \}
\end{align}
\noindent represent the sets of channel states and transmission states in the network respectively. Note that the cardinality of $\mathcal{H}$ is $\frac{|\mathcal{N}|\left(|\mathcal{N}|-1\right)}{2}$ if self-channels are disallowed and channel gains are assumed to be path symmetric. Since the cardinality of $\mathcal{X}$ is $\mathcal{|N|}$, the cardinality of $\mathsf{NSI}$ grows with the cube of the number of nodes in the network, or $O(|\mathcal{N}|^3)$. Given a half-duplex constraint, a node $i$ can either be transmitting or receiving at any given point in time\footnote{We limit the discussion to devices that can only transmit and receive. Our approach can easily be employed to consider applications such as sensor networks where devices may have additional states such as being idle.}, or $X_i \in \{ \mathsf{Tx,Rx} \}$. Let $\mathsf{R}\in \mathcal{N}$ represent a node in the network that is capable of acting as a cooperative relay for a flow of traffic in the same network and $X_\mathsf{R}$ represent its instantaneous transmission state. Additionally, let $\mathsf{\widehat{NSI}} \subset \mathsf{NSI}$ represent a subset of network state information available to the relay and $\psi \in \mathsf{\widehat{NSI}}$ represent a particular network state from the perspective of the relay. We define a {\it relaying policy} as the mapping of a known $\mathsf{NSI}$ state at $\mathsf{R}$ onto the transmission state of the relay, or
\begin{equation}
X_\mathsf{R} := f\left(\psi \right),
\end{equation}
\noindent where $f: \widehat{\mathsf{NSI}} \rightarrow \{\mathsf{Tx,Rx} \}$. We distill the task of cooperative policy design down to determining this functional mapping for a particular objective: to maximize the rate of a cooperative flow while minimizing any rate degradations in noncooperative flows of the network. In other words, we aim to minimize the spatial reuse degradation that can be caused by cooperative relays by eliminating relay transmissions in cases where doing so would harm a nearby flow.

\section{Binary Approximation and Policy Design}
\label{sct:policy}
In the model described in Section~\ref{sct:model}, $\mathsf{NSI}$ contains channel fades that are supported over a continuum of values. In this section, we develop a binary model, $\mathsf{NSI}^B$, as an approximation of the full $\mathsf{NSI}$, for the five node, two-flow network in Figure~\ref{fig:tradeoff}. We show that the states in this model can be explicitly classified by the effect that relay transmission {\it would} have on the network if the relay were to transmit in such states. We then define relaying policies that operate with incomplete $\widehat{\mathsf{NSI}}^B \subset \mathsf{NSI}^B$ and evaluate their relative performances using the binary approximation.

\subsection{Network Model Approximation}
We approximate the signal and detection models presented in Section~\ref{sct:model} in two fundamental ways. First, we consider a binary approximation of instantaneous channel fades where $H^B_{ij} \in \{0,1\}$ is a Bernoulli random variable with parameter $p_{H^B_{ij}}$. In effect, $p_{H^B_{ij}}$ acts as a proxy for $\mathsf{SNR}_{ij}$ where high $\mathsf{SNR}_{ij}$ ($p_{H^B_{ij}}\rightarrow1$) makes a high gain channel ($H^B_{ij} = 1$) very likely. Additionally, node state $X_i\in \{0,1\}$ is a Bernoulli random variable with parameter $p_{X_i}$ where a $X_i=0$ represents reception and $X_i=1$ represents transmission.

\begin{figure}[h!]
	\centering
	\includegraphics[width=3.0in]{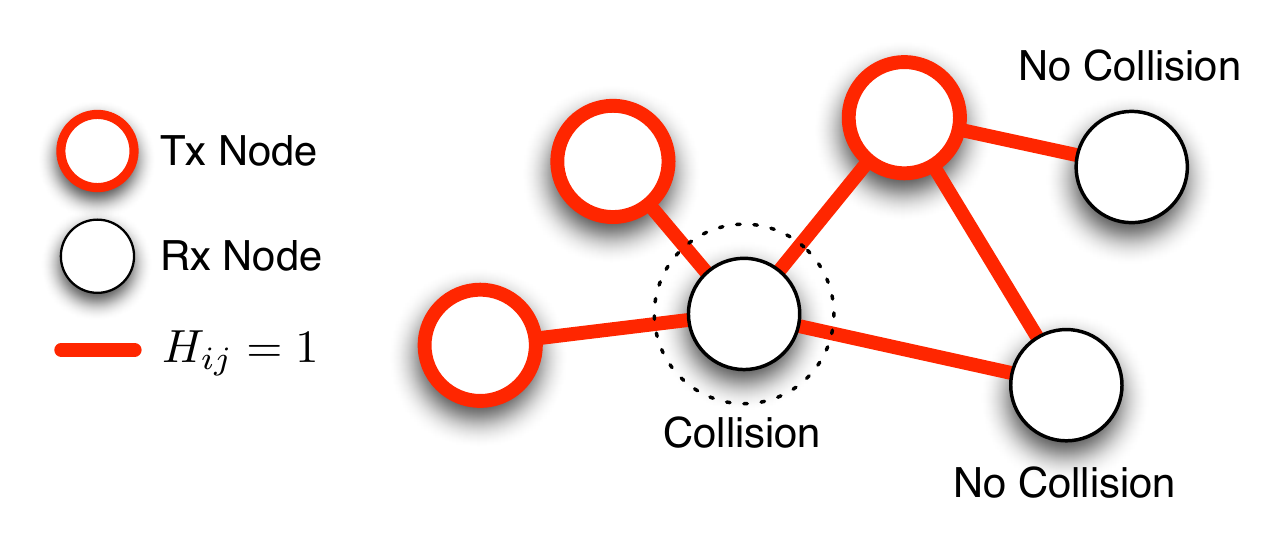}
	\caption{Nodes form vertices and channel fades form edges in the network graph.}
	\label{fig:graph}
\end{figure}

Second, we approximate the $\mathsf{SINR}$-based detection model in Section~\ref{sct:model} with a graph-based collision model illustrated in Figure~\ref{fig:graph}. In this model, nodes form vertices that are interconnected by the instantaneous edges formed by $H^B_{ij}$. If two nodes $m$ and $n$ are both in transmit states $X_m = X_n = 1$, and are linked to a common receiver $l$ with $H^B_{ml}=H^B_{nl}=1$, then a collision occurs and neither transmission is decodable. We note that the binary collision model, without the probability law on the links, is commonly used in medium access layer protocol design~\cite{tobagi2002packet}.

In Section~\ref{sct:protocols}, we remove both of these assumptions and translate the policies generated using the binary approximation into $\mathsf{SINR}$-based cooperative protocols.

\begin{figure}%
\begin{center}
    \captionsetup[figure]{margin=10pt}%
    \subfloat[15 network state elements.\label{fig:states12}]
    {\includegraphics[width=1.5in]{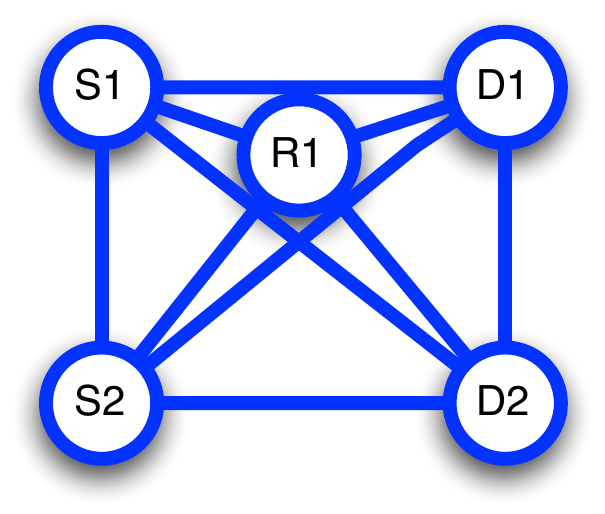}}
     \qquad
    \subfloat[8 network state elements.\label{fig:states6}]%
     {\includegraphics[width=1.5in]{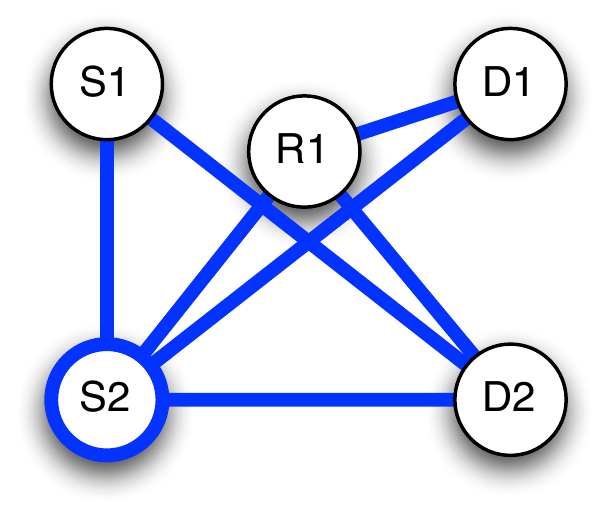}}
    \caption{The highlighted links and nodes represent network elements that can take on active or inactive states}%
    \label{fig:model}%
\end{center}
\vspace{-1.2cm}
\end{figure}

The shift to binary-valued network states reduces the uncountably infinite number of states that make up $\mathsf{NSI}$ to a finite number. That said, the cardinality of $\mathsf{NSI}^B$ still grows with the cube of the number of nodes in the network just like its continuous-valued counterpart in Equation~(\ref{eq:continuousNSI}). For tractability, we limit our study to the case of the two-flow, five node network graph shown in Figure~\ref{fig:model}. The node $\mathsf{R1}$ represents a relay node that is {\it a priori} paired with source $\mathsf{S1}$. Figure~\ref{fig:model}\subref{fig:states12} shows that 15 possible random variables (10 channel states $+$ 5 node states) make up any given snapshot of the network. Since each of these 15 bits can take on one of two values, there are a total of $2^{15}=32768$ possible network states. To reduce this state space to a more manageable size, we limit the scope of the discussion to relay policies designed for the NACK-based cooperative protocols discussed in Section~\ref{sct:model}. This reduction allows us to focus on a {\it relay-centric} network model that ignores all interactions that are unaffected by relay activity. 

The goal of this study is to determine the conditions under which the relay should transmit (i.e. when $X_{\mathsf{R1}}=\mathsf{Tx}$). Let $\psi^B \in \mathsf{NSI}^B$ represent a single state of the network. This state is formed by the $H^B_{ij\in\mathcal{N}}$ and $X_{i\in\mathcal{N}}$ bits present in the two-flow network. We need not consider the value of $X_{\mathsf{R1}}$ in the construction of $\psi^B$ because the goal is to determine $X_{\mathsf{R1}}$ as a function of the other elements. Second, by assuming a NACK-based protocol where the relay is only ever requested to transmit under the condition that its source is unable to communicate to its destination, we can further reduce the following states as follows:

\begin{itemize}
\item{ $X_{\mathsf{S1}}\equiv1$: A NACK from $\mathsf{D1}$ triggers {\it simultaneous} transmissions at $\mathsf{S1}$ and $\mathsf{R1}$. If it decides to transmit, $\mathsf{R1}$ will overlap transmission with $\mathsf{S1}$.}
\item{ $X_{\mathsf{D1}}\equiv0$: If the cooperation request is signaled by $\mathsf{D1}$ via a NACK, then $\mathsf{D1}$ knows that a cooperative transmission is to follow and it will not initiate any transmissions.}
\item{ $X_{\mathsf{D2}}\equiv0$: In general, $\mathsf{D2}$ can potentially generate transmissions in the form of ACK/NACK control packets meant for $\mathsf{S2}$. To reduce the number of states that must be considered, we assume that this cannot occur. In Section~\ref{sct:protocols}, we broaden the policies generated by this model to include arbitrary number of flows among an arbitrary number of nodes. Since flows can be bidirectional, this effectively also captures the case of interference caused by control packets and thus relaxes this assumption.}
\item{ $H^B_\mathsf{S1D1}\equiv0$: In a reactive cooperative protocol, relay transmissions only occur when the corresponding source transmission fails due to insufficient channel gain. Thus, we can assume that this channel is disconnected\footnote{A relaying phase is triggered by an explicit NACK broadcast from destination to source and relay. Hence, an assumption that $H^B_\mathsf{S1D1}=0$ appears dissatisfying since the NACK must be communicated over this channel back to the source. In practice, NACKs can be coded at far lower rates and thus be far more resilient to channel outages than data payloads. Thus, $H^B_\mathsf{S1D1}=0$ represents the case the channel gain is low enough to not support a full data payload yet high enough to support a NACK.}.}
\item{ $H^B_\mathsf{S1R1}\equiv1$: Given the decode-and-forward physical layer operating at the relay, the link between $\mathsf{S1}$ and $\mathsf{R1}$ must be connected for the relay to transmit.}
\end{itemize}

Figure~\ref{fig:model}\subref{fig:states6} shows that this conditional model reduces the number of state elements in the network to only 8, leaving a far more manageable total of $2^8=256$ possible states.

\subsection{State Classification}
A relay transmission can have a number of effects on the network as a whole. We classify these effects into three sets $\mathcal{A}$, $\mathcal{B}$, and $\mathcal{C}$. Set $\mathcal{A}$ contains all states where a relay transmission {\it assists} the $\mathsf{S1}$-$\mathsf{D1}$ flow in recovering a packet. Set $\mathcal{B}$ contains all states where $\mathsf{S2}$ is forced to defer a {\it backoff} while receiving when it otherwise would not because of $\mathsf{R1}$ triggering a carrier-sense. Finally, set $\mathcal{C}$ contains all states where $\mathsf{D2}$ fails to decode a message from $\mathsf{S2}$ because of a {\it collision} caused by $\mathsf{R1}$. Formally,
\begin{align}
\label{eq:setA}
\mathcal{A} &\in \{\psi^B | H^B_\mathsf{R1D1} \overline{X_{\mathsf{S2}} H^B_\mathsf{S2D1}}=1 \} \\
\label{eq:setB}
\mathcal{B} &\in \{\psi^B |  \overline{X_\mathsf{S2}} H^B_\mathsf{R1S2} \overline{H^B_\mathsf{S1S2}} =1\} \\
\label{eq:setC}
\mathcal{C} &\in \{\psi^B |  X_\mathsf{S2} H^B_\mathsf{S2D2} H^B_\mathsf{R1D2} \overline{H^B_\mathsf{S1D2}} = 1  \}
\end{align}
\noindent where the overline represents a logical complement.

\begin{figure}%
\begin{center}
    \captionsetup[figure]{margin=10pt}%
    \subfloat[Assist Example ($\mathcal{A}$).\label{fig:help}]
    {\includegraphics[width=1.5in]{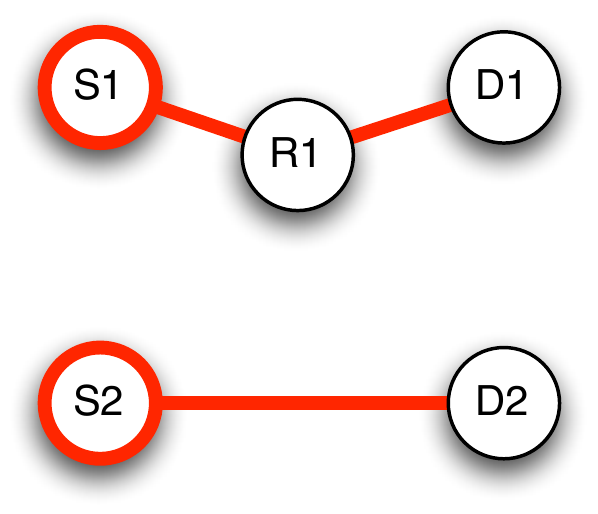}}
     \qquad
    \subfloat[Backoff Example ($\mathcal{B}$).\label{fig:defer}]
    {\includegraphics[width=1.5in]{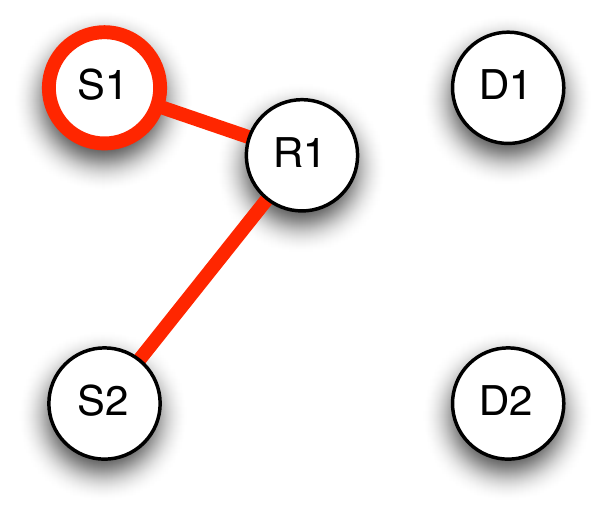}}
    \qquad
         \subfloat[Collision Example ($\mathcal{C}$).\label{fig:drop}]
    {\includegraphics[width=1.5in]{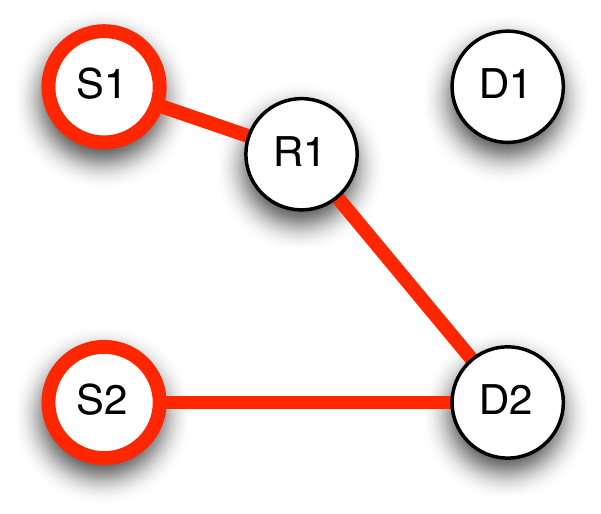}}
    \caption{Every state can be labelled with membership in the $\mathcal{A}$, $\mathcal{B}$, and $\mathcal{C}$ subsets.}%
    \label{fig:examples}%
\end{center}
\vspace{-.6cm}
\end{figure}

Figure~\ref{fig:examples} highlights three example network states $\psi^B$ that occupy the subsets $\mathcal{A}$, $\mathcal{B}$, and $\mathcal{C}$. Since events $\mathcal{B}$ and $\mathcal{C}$ correspond to mutually exclusive events  ($\mathsf{S2}$ reception and transmission respectively), these subsets are also mutually exclusive. Using Equations~(\ref{eq:setA}) through (\ref{eq:setC}), we label each network state $\psi^B$ with its membership in these subsets, or
\begin{equation}
\label{eq:statelabels}
\psi^B \in \{\mathcal{A},\mathcal{B}, \mathcal{C}, \mathcal{A}\cap \mathcal{B}, \mathcal{A} \cap \mathcal{C},\mathcal{D} \},
\end{equation}

\noindent where $\mathcal{D}$ represents a set of states where relay transmission has neither positive nor negative impact on the network. In Appendix~\ref{sct:appendix}, we classify each possible network state.

\subsection{Relaying Policies with Partial NSI}

In the previous section, we showed that network states can be classified according to the relay's effects on the network. Given these labels, relaying policies can be derived that govern whether a relay should transmit as a function of the current state of the network ($X_\mathsf{R1}=f \left(\psi^B \right)$). In this section, we first define relaying policies assuming that the relay is fully aware of the current global network state $\psi^B$. We then consider relaying policies where the relay has incomplete network state information ($\widehat{\mathsf{NSI}}^B$).

\subsubsection{Full NSI}
\label{sct:fullNSI}

When a relay has access to full $\mathsf{NSI}^B$, it can accurately determine the current state of the network $\psi^B$. As such, the relay knows perfectly what effect transmission during this state will have on the network as a whole. We can define a relay policy that minimizes negative impact on a surrounding flow by disallowing transmission when $\psi^B$ is labelled with events $\mathcal{B}$ or $\mathcal{C}$ since these reduce spatial reuse by interfering with the operations of the other flow:
\begin{align}
X_\mathsf{R1} &=
  \begin{cases}
   0 & \text{if } \psi^B \in \mathcal{B}\cup\mathcal{C} \\
   1 & \text{if } \psi^B \in \mathcal{A}\cap \left(\overline{\mathcal{B}\cup\mathcal{C}}\right) \\
   Z & \text{otherwise},
  \end{cases}
\end{align}
\noindent where $Z$ represents a ``don't care'' where neither a relay transmission nor the lack thereof will impact the network in any way. Counting the number of states that are members of $\mathcal{B}$ or $\mathcal{C}$ in Appendix~\ref{sct:appendix}, we see that 48 of the 256 total states represent conditions where the relay should avoid transmitting. One can write the Boolean expression that ties the values of the individual network state elements to the behavior of the relay $\overline{X_\mathsf{R1}}$ (the relay avoiding transmission). One can employ standard Boolean reduction techniques to convey this behavior more simply than the sum-of-products form of 48 cases, or
\begin{equation}
\label{eq:fullNSI}
\overline{X^{\mathsf{FNSI}}_\mathsf{R1}} = \underbrace{ X_\mathsf{S2} H^B_\mathsf{S2D2} H^B_\mathsf{R1D2}\overline{H^B_\mathsf{S1D2}}}_{\text{\ding{172}}}+ \underbrace{\overline{X_\mathsf{S2}}H^B_\mathsf{R1S2}\overline{H^B_\mathsf{S1S2}}}_{\text{\ding{173}}}
\end{equation}
In this expression, we use the $\mathsf{FNSI}$ acronym as representation of the ``Full NSI'' policy. There are two critical components to this behavior. The \ding{172} term addresses the potential for the relay to cause a packet drop due to a collision with a transmission from $\mathsf{S2}$. Specifically, the relay should avoid transmitting when $\mathsf{S2}$ is transmitting and $\mathsf{S2}$ would not collide with a transmission from $\mathsf{S1}$ but would collide with a transmission from $\mathsf{R1}$. The \ding{173} term addresses the potential for the relay to cause unnecessary backoff deferrals at $\mathsf{S2}$. The relay should avoid transmitting when $\mathsf{S2}$ is receiving and no link is present between $\mathsf{S1}$ and $\mathsf{S2}$ but a link is present between $\mathsf{S1}$ and $\mathsf{R1}$. This behavior establishes the baseline performance of a relaying policy that has access to all of the elements required to calculate Equation~(\ref{eq:fullNSI}). The power of this methodology lies in the fact that we can also determine the relay behavior for any arbitrary subset of $\mathsf{NSI}^B$.

\subsubsection{Incomplete NSI}
\label{sct:incompleteKnowledge}
One can use exactly the same full $\mathsf{NSI}^B$ table in Appendix~\ref{sct:appendix} to construct incomplete $\mathsf{NSI}^B$ policies by recognizing that eliminating knowledge is equivalent to binning network states into coarser delineation.

\begin{figure}%
\begin{center}
    \captionsetup[figure]{margin=10pt}%
    \subfloat[$\overline{X^{\mathsf{FNSI}}_\mathsf{R1}} = 0$ since $\psi^B \in \mathcal{A}$. \label{fig:off}]
    {\includegraphics[width=1.5in]{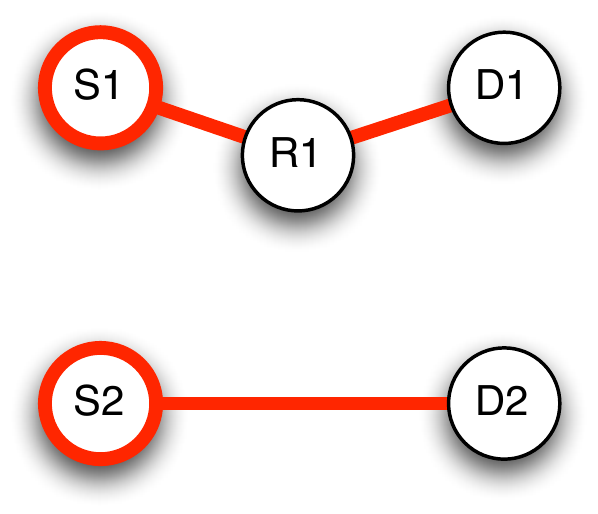}}
     \qquad
    \subfloat[$\overline{X^{\mathsf{FNSI}}_\mathsf{R1}} = 1$ since $\psi^B \in \mathcal{C}$. \label{fig:on}]%
     {\includegraphics[width=1.5in]{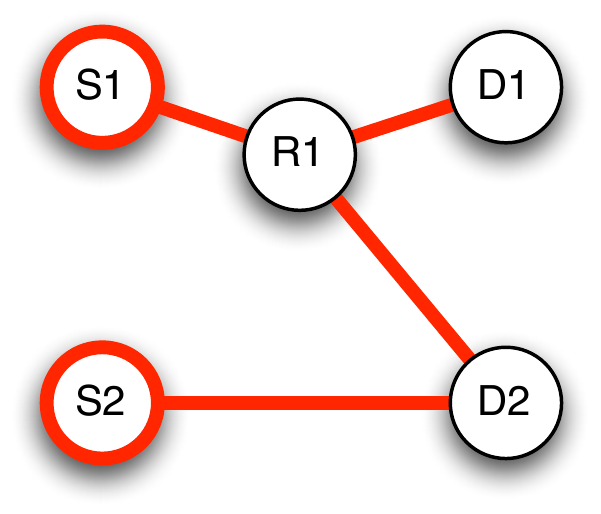}}
    \caption{If $H_\mathsf{R1D2}$ is unknown, the relay cannot distinguish between these two states.}%
    \label{fig:know}%
\end{center}
\vspace{-.6cm}
\end{figure}

Figures~\ref{fig:know}\subref{fig:off} and \ref{fig:know}\subref{fig:on} show two network states where the full $\mathsf{NSI}$ policy enables and disables relay transmission respectively. However, the only difference between the network states is the $H_\mathsf{R1D2}$ state element. If this state element is unknown to the relay, the two states are binned together creating a {\it conflict set} where the lack of knowledge yields ambiguity in what the relay should do; the relay is unable to determine whether the current state of the network is in an assist classification $\mathcal{A}$ or also in a collision classification $\mathcal{C}$. In dealing with these conflict sets that arise with incomplete knowledge available to the relay, we consider two approaches to this problem:

\noindent {\bf Conservative View:} When a relay is unable to distinguish between multiple states, it may assume the worst about the state elements it does not know. This assumption yields a disabled relay ($X_\mathsf{R1}=0$) in the case that {\it any} state within the conflict set demands a disabled relay.

\noindent {\bf Greedy View:} Adopting the best-case viewpoint about unknown states, a relay can enable transmission ($X_\mathsf{R1}=1$) when {\it any} state within the conflict set demands relay transmission.

These approaches apply to any arbitrary subset of the full $\mathsf{NSI}^B$ knowledge, so a remaining task is to determine what subsets of full $\mathsf{NSI}^B$ to consider. A useful way of sorting $\mathsf{NSI}^B$ is considering the hop-distance of the $\mathsf{NSI}^B$ elements from the relay. This approach allows a quantitative description of how ``local'' a node's view of the network is~\cite{aggarwal2009sum}.

 Let $\widehat{\mathsf{NSI}}^B\left(n\right)$ represent the set of $\mathsf{NSI}$ elements no further than $n$ hops away from the relay. In our two-flow network, these sets are defined as
\begin{align}
\widehat{\mathsf{NSI}}^B\left(2\right) &= \{ H^B_ \mathsf{S1S2}, H^B_ \mathsf{R1S2}, H^B_ \mathsf{S1D2},H^B_ \mathsf{S2D2}, H^B_ \mathsf{R1D2}, H^B_ \mathsf{R1D1},H^B_ \mathsf{S2D1} \} \\
\widehat{\mathsf{NSI}}^B\left(1\right) &= \{ H^B_ \mathsf{R1S2} , H^B_ \mathsf{R1D2}, H^B_ \mathsf{R1D1} \} \\
\widehat{\mathsf{NSI}}^B\left(0\right) &= \{ \emptyset \}.
\end{align}
In the case of $\widehat{\mathsf{NSI}}^B\left(2\right)$, all wireless channels are at least two hops away so the relay knows full $\mathsf {NSI}^B$ with the notable exception of the  transmission state of $\mathsf{S2}$ (i.e. ${X_\mathsf{S2}}$).

In the following sections we use the notation $\mathsf{Cons}(n)$ to identify policies that use the conservative mapping with $n$ hops of knowledge. Similarly, the notation $\mathsf{Greed}(n)$ is used to identify policies that use the greedy mapping.

\noindent {\bf Conservative Policies:} 

Using the same Boolean reduction techniques as before, conservative relaying policies can be identified for different numbers of hops of information made available to the relay.
\begin{align}
\label{eq:cons2}
\overline{X^{\mathsf{Cons}(2)}_\mathsf{R1}} &= H^B_\mathsf{S2D2} H^B_\mathsf{R1D2}\overline{H^B_\mathsf{S1D2}}  +H^B_\mathsf{R1S2}\overline{H^B_\mathsf{S1S2}}\\
\label{eq:cons1}
\overline{X^{\mathsf{Cons}(1)}_\mathsf{R1}} &=  H^B_ \mathsf{R1S2}+H^B_ \mathsf{R1D2}\\
\label{eq:cons0}
\overline{X^{\mathsf{Cons}(0)}_\mathsf{R1}} &=  1.
\end{align}
In the full $\mathsf{NSI}^B$ case in Equation~(\ref{eq:fullNSI}), the $X_ \mathsf{S2}$ acts as a kind of switch to determine whether the relay's behavior is dominated by collision avoidance or backoff deferral avoidance. In Equation~(\ref{eq:cons2}), this switch is missing and both terms apply because the two-hop policy does not have access to the node state. In Equation~(\ref{eq:cons1}), the relay is only able to base its decision of whether to transmit on the set $\widehat{\mathsf{NSI}}^B\left(1\right)$. Acting conservatively, the relay is only able to transmit when the links between $\mathsf{R1}$ and {\it both} $\mathsf{S2}$ and $\mathsf{D2}$ are disconnected. The relay guarantees that it cannot cause a backoff deferral at $\mathsf{S2}$ or a collision at $\mathsf{D2}$. Finally, the relay in Equation~(\ref{eq:cons0}) is never able to transmit since it can never guarantee that it will not harm another flow.

\noindent {\bf Greedy Policies:} 

Greedy relaying policies can be identified for different numbers of hops of information made available to the relay.
\begin{align}
\label{eq:greed2}
\overline{X^{\mathsf{Greed}(2)}_\mathsf{R1}} &=   H^B_\mathsf{S2D2} H^B_\mathsf{R1D2}\overline{H^B_\mathsf{S1D2}}  \cdot H^B_\mathsf{R1S2}\overline{H^B_\mathsf{S1S2}}+ \overline{H^B_\mathsf{R1D1}}\\
\label{eq:greed1}
\overline{X^{\mathsf{Greed}(1)}_\mathsf{R1}} &=  \overline{H^B_\mathsf{R1D1}}\\
\label{eq:greed0}
\overline{X^{\mathsf{Greed}(0)}_\mathsf{R1}} &=  0.
\end{align}
The key difference between Equations~(\ref{eq:fullNSI}) and (\ref{eq:greed2}) is that if only one condition (collision or backoff deferral) instructs the relay to halt, it is assumed that the unknown $X_\mathsf{S2}$ state element would have disabled that term. In other words, the relay only halts transmissions when either a $\mathcal{B}$ or $\mathcal{C}$ event would occur regardless of the $X_\mathsf{S2}$ state. Additionally, another case for disabling the relay appears when $\mathsf{R1}$ and $\mathsf{D1}$ are disconnected since the relay cannot assist the cooperative flow in this case. In Equation~(\ref{eq:greed1}), the relay only disables transmission when it knows that it will not be able to help. In these cases, there is {\it only} an opportunity to harm the network, so even the greedy relay disables transmission. Finally, the relay in Equation~(\ref{eq:greed0}) knows nothing about the network and greedily transmits whenever it is requested.

\subsection{Discussion of Protocol Overhead}
The binary model-based relay policies dictate the behavior of the relay \emph{given} elements of NSI. In this section, we discuss how such information might be collected in actual protocols, noting that complete protocol implementation is out of scope of this paper. The amount of overhead for collecting this information is determined by two factors: (i)~the rate of change of $\mathsf{NSI}$ and (ii)~how much knowledge is desired. 

First, the rate of change of $\mathsf{NSI}$ depends on the amount of mobility in the system. For low-mobility, slow-fading environments such as the indoor Wi-Fi, channel coherence times can be many tens or hundreds of packet intervals. As such, $\mathsf{NSI}$ knowledge at the relay need only be updated on the timescales of these coherence times. Second, the amount of overhead required differs from one $\mathsf{NSI}$ element to the next. For instance, the one-hop $\mathsf{NSI}$ states may be logged passively (with zero overhead) at the relay by simply overhearing surrounding transmissions. In fact, even some two-hop knowledge may be acquired without additional overhead. Assuming the non-cooperative flow employs the same NACK-based protocol as the cooperative flow, the relay can infer the link quality between the non-cooperative source and destination by overhearing ACKs and NACKs generated by the non-cooperative destination. 

\subsection{Performance Evaluation}

As a mechanism to compare the performance of different policies, we evaluate the probability of the network entering a particular event subset while simultaneously considering whether a relaying policy transmits. In other words, a relaying policy can be penalized for transmitting within the $\mathcal{B}$ or $\mathcal{C}$ event subsets and rewarded for transmitting in $\mathcal{A}$. The probability of a relay transmitting in event $\mathcal{A}$ is 
\begin{align}
\label{eq:performance}
Pr\{X_\mathsf{R1}& \cap \psi^B \in \mathcal{A}\} = \sum_{\psi^B\in\mathcal{A}} X_\mathsf{R1}\left(\psi^B\right) \cdot Pr\{\psi^B \},
\end{align}
\noindent where $Pr\{\psi^B\}$ can be calculated by the product of the Bernoulli parameters. Similar expressions for event spaces $\mathcal{B}$ and $\mathcal{C}$ can be derived.

It is useful to consider a particular application scenario where the $\mathsf{R1}$ node is geographically near the $\mathsf{S1}$ node. This models a usage case where one user owns both the relay and source nodes and both devices are located near the user. Furthermore, let us simplify the discussion of these systems by considering a single dominant parameter: flow separation. Specifically, let $p_{H^B_\mathsf{S1S2}}=p_{H^B_\mathsf{S1D2}}=p_{H^B_\mathsf{R1S2}}=p_{H^B_\mathsf{R1D2}}=p$ where $p$ represents a single parameter that acts as a proxy for flow separation. As $p\rightarrow1$, the flows are topologically connected with high probability, and as $p\rightarrow0$, the flows are disconnected with high probability. For simplicity of discussion, assume every other state element probability is $\frac{1}{2}$. Using Equation~(\ref{eq:performance}), we compute expressions that determine the propensity of each policy to transmit in the $\mathcal{A}$, $\mathcal{B}$, and $\mathcal{C}$ subsets as a function of the single independent parameter $p$.

\begin{table}[h!]
\caption{Performance Evaluation of Relaying Policies}
\begin{center} 
\begin{tabular}{|c||c|c|c|}
\hline
Policy & $Pr\{X_\mathsf{R1} \cap \psi^B \in \mathcal{A}\}$ & $Pr\{X_\mathsf{R1} \cap \psi^B \in \mathcal{B}\}$ & $Pr\{X_\mathsf{R1} \cap \psi^B \in \mathcal{C}\}$\\
\hline
\hline
$\mathsf{FNSI}$ & $\frac{5p^2-5p+6}{16}$ & 0 & 0 \\
\hline
\hline
$\mathsf{Cons}(2)$ & $\frac{3p^4-6p^3+10p^2-7p+2}{16}$ & 0 & 0 \\
\hline
$\mathsf{Cons}(1)$ & $\frac{3p^2-6p+3}{8}$ & 0 & 0 \\
\hline
$\mathsf{Cons}(0)$ & 0 & 0 & 0 \\
\hline
\hline
$\mathsf{Greed}(2)$ & $\frac{-p^4+2p^3-p+3}{8}$ & $\frac{-p^4+2p^3-3p^2+2p}{16}$ & $\frac{-p^4+2p^3-2p^2+p}{8}$ \\
\hline
$\mathsf{Greed}(1)$ & $\frac{3}{8}$ & $\frac{-p^2+p}{4}$ & $\frac{-p^2+p}{8}$ \\
\hline
$\mathsf{Greed}(0)$ & $\frac{3}{8}$ & $\frac{-p^2+p}{2}$ & $\frac{-p^2+p}{4}$ \\
\hline
\end{tabular}
\label{tbl:performance}
\end{center}
\end{table}

Table~\ref{tbl:performance} summarizes the performance of each of the six previously described relaying policies. Of note, the full $\mathsf{NSI}^B$ policy and the conservative incomplete $\mathsf{NSI}^B$ policies never transmit in the subsets where doing so could potentially cause a collision or backoff deferral event. As such, the probability of harming the other flow by transmitting on these occasions is zero. The greedy policies, however, allow some degradation in the other flow in order to improve the policies' abilities to assist their own flows. 

\begin{figure}[h!]%
\begin{center}
    \captionsetup[figure]{margin=10pt}%
    \subfloat[Probability of assisting.\label{fig:help_hop}]
    {\includegraphics[width=1.9in]{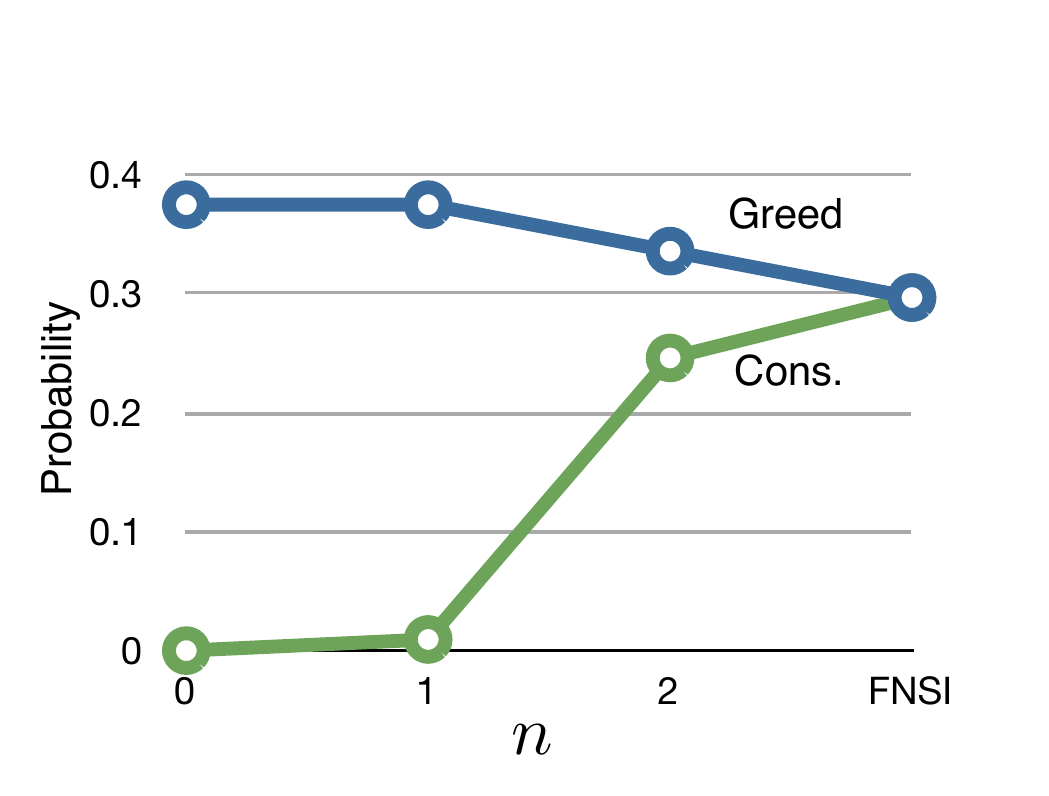}}
     \qquad
         \subfloat[Probability of causing backoff deferral.\label{fig:defer_hop}]
    {\includegraphics[width=1.9in]{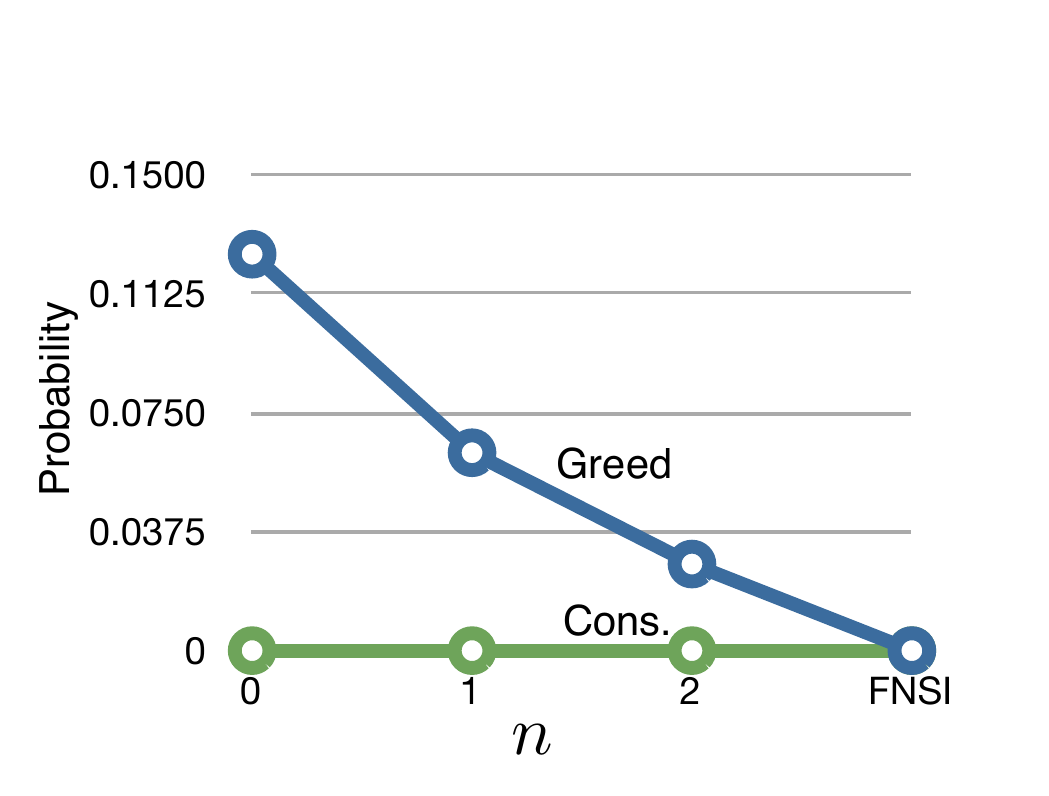}}
     \qquad
    \subfloat[Probability of causing collision.\label{fig:drop_hop}]
    {\includegraphics[width=1.9in]{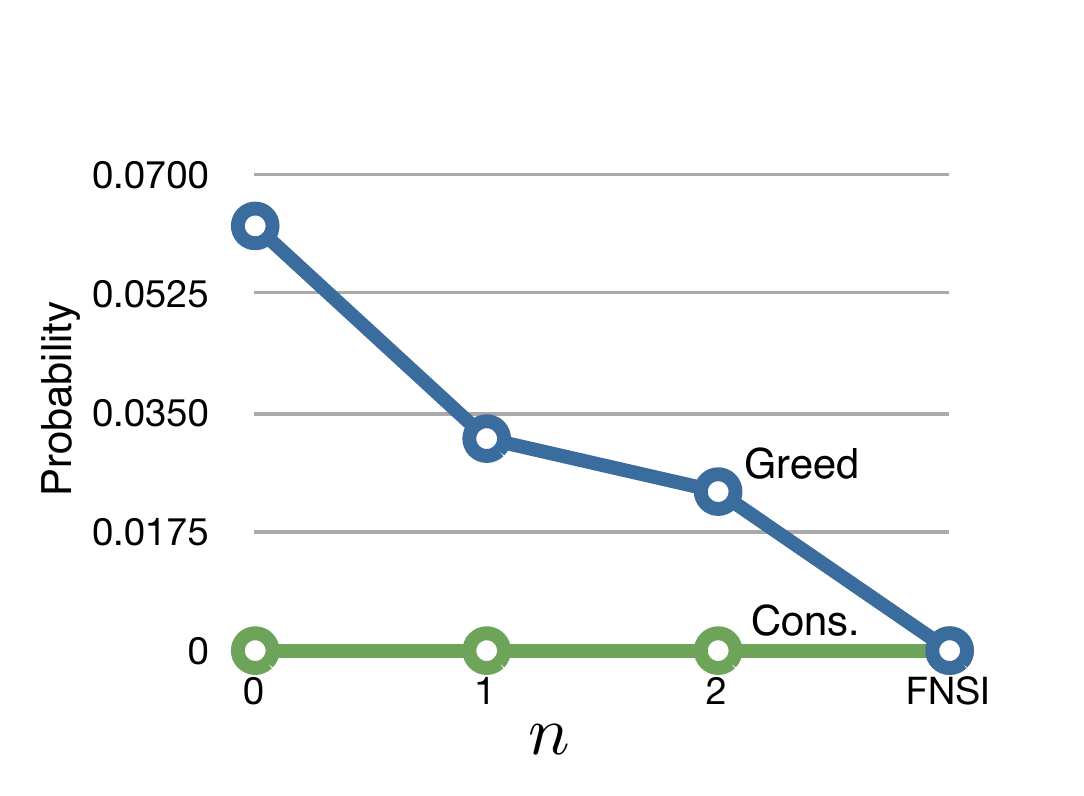}}
    \caption{Conservative relaying behavior requires substantial $\mathsf{NSI}$ before cooperative gain is observed.}%
    \label{fig:policyPerformanceHop}%
\end{center}
\end{figure}

Consider the case that $p=\frac{1}{2}$. Figure~\ref{fig:policyPerformanceHop} shows the performance of each policy as a function of $\mathsf{NSI}^B$ available to $\mathsf{R1}$. In general, the trend is that more $\widehat{\mathsf{NSI}}^B$ knowledge results in less harm to another flow since the relay knows more about the network it needs to protect. Likewise, increasing $\widehat{\mathsf{NSI}}^B$ knowledge allows conservative relays to assist their flow more and eventually converge with their greedy counterparts. Incrementally, the jump from zero hops of knowledge to one hop of knowledge has very little effect on the conservative policies---the improvement seen in performance of the cooperative flow is marginal. For the greedy policies, however, having even a single hop of information provides a substantial drop in the amount of harm the relay will impart on the neighboring flow. Conservative policies require large amounts of $\mathsf{NSI}$ before cooperative gains can be seen.

\begin{figure}[h!]%
\begin{center}
    \captionsetup[figure]{margin=10pt}%
    \subfloat[Probability of assisting.\label{fig:help_all}]
    {\includegraphics[width=1.9in]{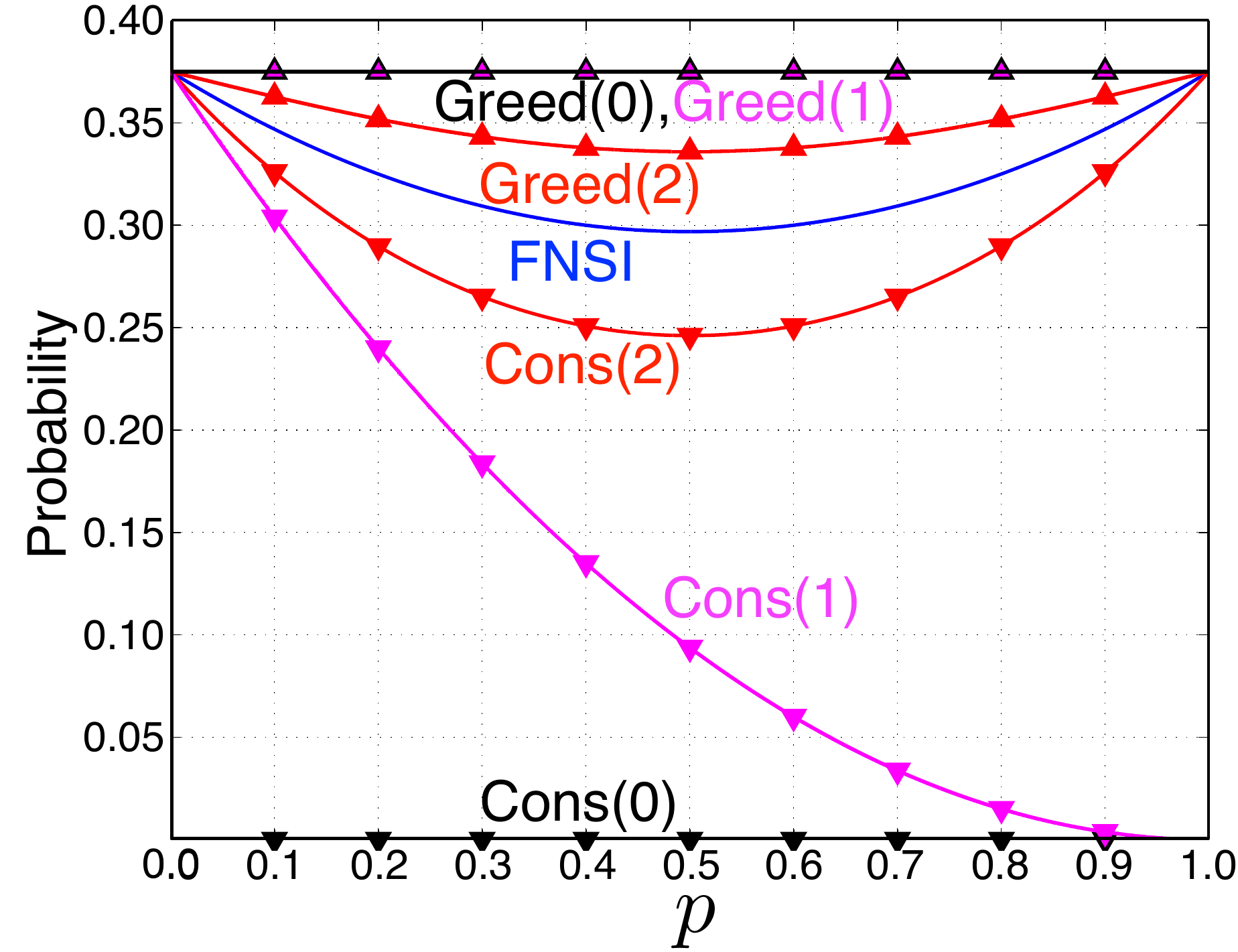}}
     \qquad
         \subfloat[Probability of causing backoff deferral.\label{fig:defer_all}]
    {\includegraphics[width=1.9in]{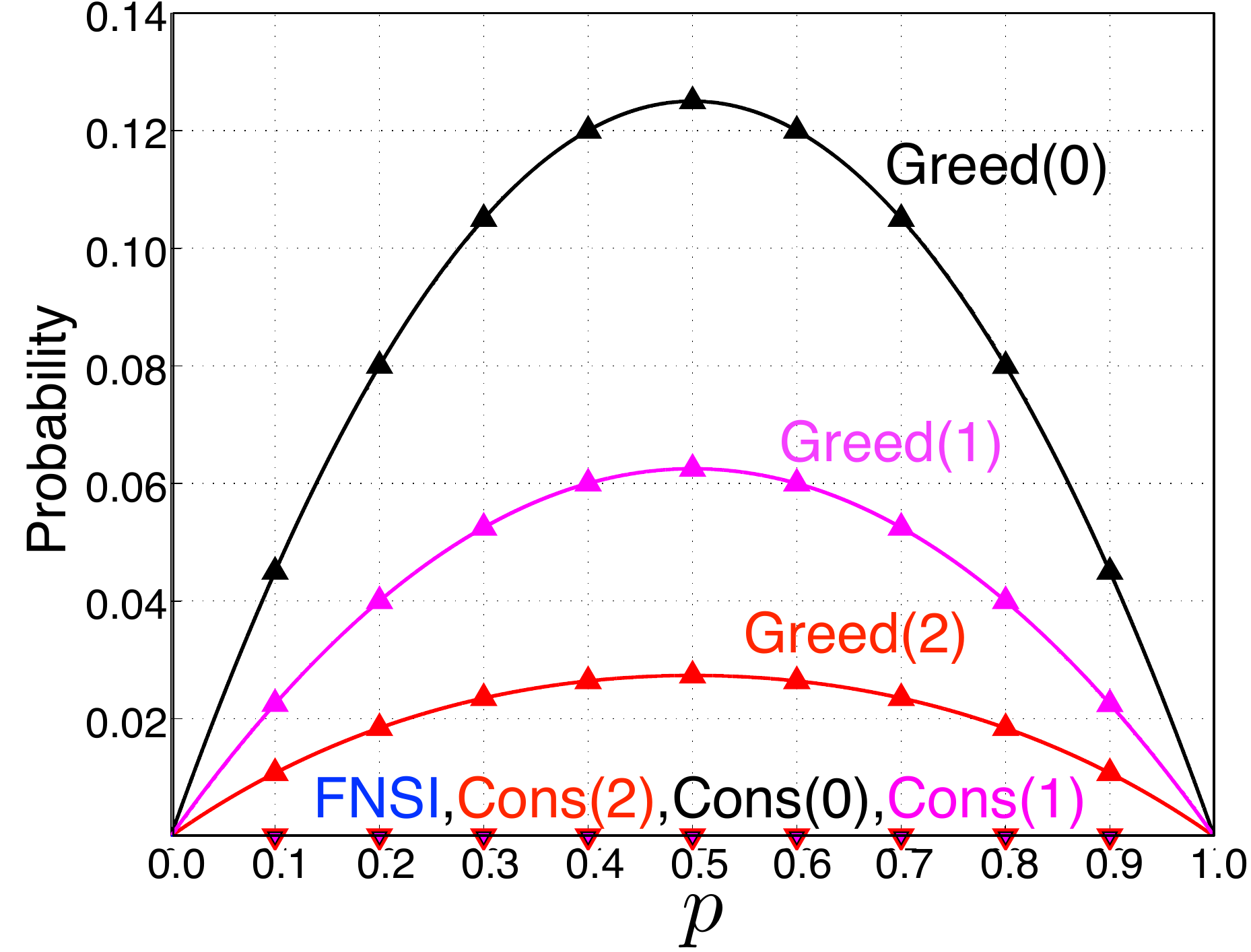}}
     \qquad
    \subfloat[Probability of causing collision.\label{fig:drop_all}]
    {\includegraphics[width=1.9in]{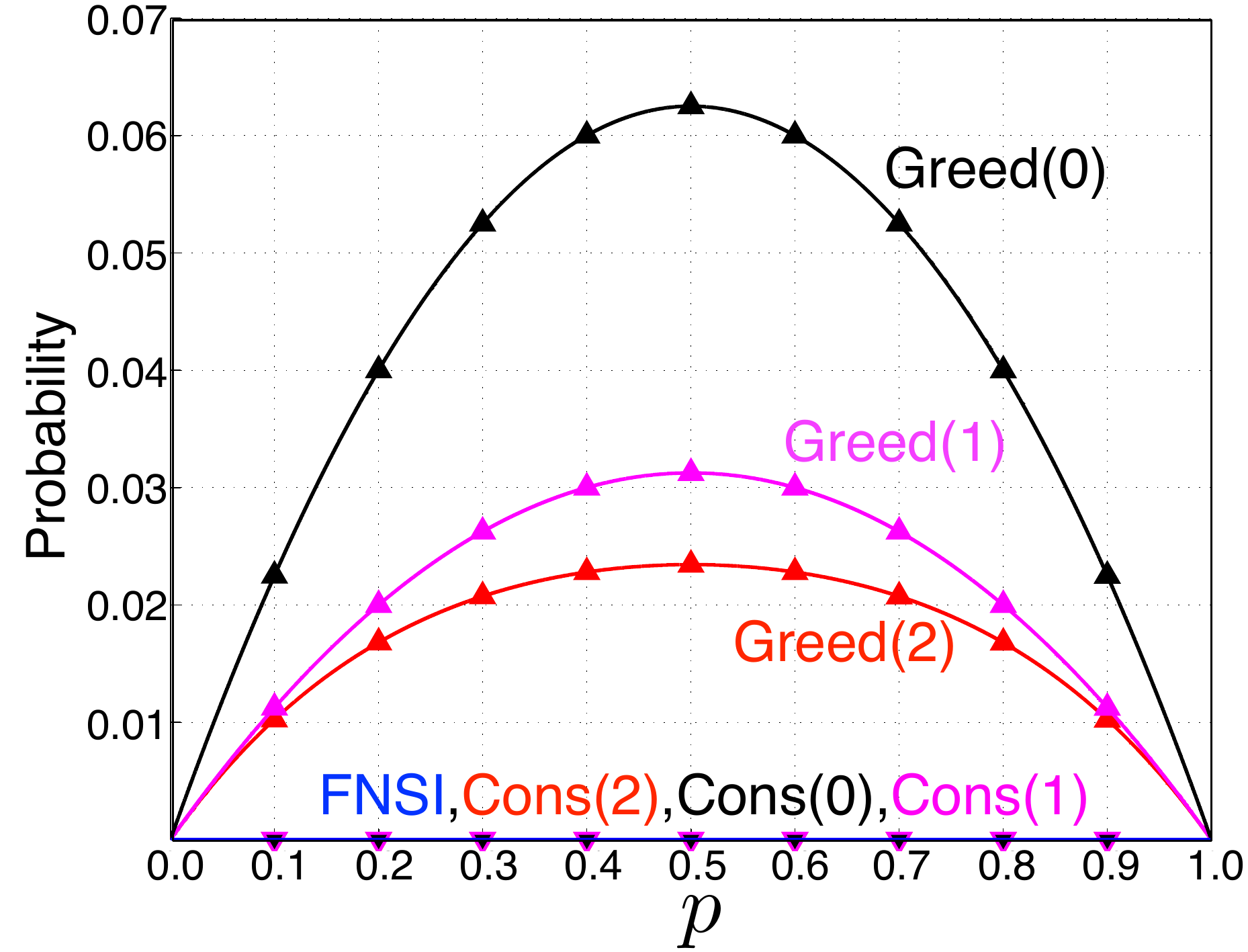}}
    \caption{Each policy exhibits different behaviors in terms of the relay's propensity to transmit in the event subspaces.}%
    \label{fig:policyPerformance}%
\end{center}
\end{figure}

Figure~\ref{fig:policyPerformance} plots the expressions in Table~\ref{tbl:performance} as functions of $p$. In Figure~\ref{fig:policyPerformance}\subref{fig:help_all}, we plot the probability of each scheme transmitting during the states where a relay is able to help its paired flow. The greedy policies all improve performance over the full $\mathsf{NSI}^B$ policy since they transmit during cases where the full $\mathsf{NSI}^B$ policy halts relay transmission in accordance with minimizing negative impact on the neighboring flow. The conservative policies decrease performance over the full $\mathsf{NSI}^B$ policy since they avoid transmitting in states where the full $\mathsf{NSI}^B$ policy would because they are unable to distinguish these states from those where the relay should be halted. The $\mathsf{Cons}(1)$ policy in particular exhibits an unusual behavior in that it is able to help only as $p\rightarrow 0$. This is due to the fact that, given only one hop of $\mathsf{NSI}^B$ knowledge, a relay is unable to align its transmissions to source interference that would be present anyway since it has no idea what the link qualities are between its source and other nodes in the network.

In Figures~\ref{fig:policyPerformance}\subref{fig:defer_all} and \ref{fig:policyPerformance}\subref{fig:drop_all}, we see the impact of the relay policies on the probability of the neighboring flow deferring and colliding, respectively. As stated previously, the full $\mathsf{NSI}^B$ and conservative incomplete $\mathsf{NSI}^B$ policies cause no deferrals or collisions. The greedy incomplete $\mathsf{NSI}$ policies, however, allow degradations in the interest of increasing the probability of assisting the cooperative flow.

\section{Protocol Design and Simulation}
\label{sct:protocols}
The policies presented in Section~\ref{sct:policy} operate on binary network state information. Now, we translate the preceding two-flow policies into $n$-flow protocols. These protocols are then implemented in a custom network simulator and are evaluated in realistic fading environments.

\subsection{Protocol Translation}
The binary network model abstracts from reality in two key ways. First, only two unidirectional flows are considered, whereas arbitrary networks can potentially have many bidirectional flows. Second, channels take on only binary states whereas actual channels span a continuum of powers. We now translate the aforementioned policies into cooperative protocols that overcome these limitations of the model. 

Specifically, we can directly translate the $\mathsf{Cons}(2)$, $\mathsf{Cons}(0)$, $\mathsf{Greed}(2)$, and $\mathsf{Greed}(0)$ policies. The one-hop policies, however, highlight a limitation in the binary network model when it applies to SINR-based protocol design. Consider the policy stated in Equation~({\ref{eq:cons1}). The relay transmits when its links to the other flow are disconnected. As defined by the binary collision model, the relay is able to guarantee that no collision or deferral event can take place in these states. This policy does not translate to an $\mathsf{SINR}$-based scenario, where the measurement of a single link quality is insufficient to guarantee that a collision or deferral event will not take place. Even if the relay measured the instantaneous channel between itself and another destination as being very weak, it is still possible that a transmission will cause a collision if the channel supporting the other flow is also very weak. Despite this limitation, we are able to conclusively show that the remaining policies not only are capable of being translated into $\mathsf{SINR}$-based protocols, but their \emph{relative} performance in realistic fading environments is accurately predicted by our analysis of the binary model.

\noindent {\bf Conservative Protocols:}

The $\mathsf{Cons}(2)$ and $\mathsf{Cons}(0)$ policies can be directly translated into protocols that operate on instantaneous $\mathsf{SINR}$ measurements. Consider a network $\mathcal{N}$ consisting of $N$ nodes.

\begin{algorithm}[h!]
\label{al:cons2}
$\mathcal{N} = \{0,1,2,\ldots,N-1\}$\\
$X^{\mathsf{Cons}(2)}_\mathsf{R} = \mathsf{Tx}$\\
\For{$i \in \mathcal{N}\setminus \{\mathsf{S},\mathsf{D},\mathsf{R}\}$}{
\For{$j \in \mathcal{N}\setminus \{i,\mathsf{S},\mathsf{D},\mathsf{R}\}$}{
\If{($\overline{\mathsf{BO}^{\mathsf{S}}_i}$ and $\mathsf{BO}^{\mathsf{S} \mathsf{R}}_i$) or ($\overline{\mathsf{COL}^{\mathsf{S}}_{i j}}$ and $\mathsf{COL}^{\mathsf{S}\mathsf{R}}_{i j}$)}{$X^{\mathsf{Cons}(2)}_\mathsf{R} = \mathsf{Rx}$}
}
}
\caption{$\mathsf{Cons}(2)$}
\end{algorithm}

Protocol~\ref{al:cons2} formally specifies the $\mathsf{Cons}(2)$ behavior. The collision and backoff terms are
\begin{align}
\mathsf{COL}^{\mathsf{S}}_{ij} &=\left[\frac{P_t L_{ij} |h_{ij}|^2}{P_t L_{\mathsf{S}j}|h_{\mathsf{S}j}|^2  + z } < \gamma_\mathsf{DET}\right] \nonumber\\
\mathsf{COL}^{\mathsf{S}\mathsf{R}}_{ij} &=\left[\frac{P_t L_{ij} |h_{ij}|^2}{P_t L_{\mathsf{S}j}|h_{\mathsf{S}j}|^2 + P_t L_{\mathsf{R}j}|h_{\mathsf{R}j}|^2 + \sum_{k\in\mathcal{I}_1} P_t L_{kj}|h_{kj}|^2 + z } < \gamma_\mathsf{DET}\right] \nonumber\\
\mathsf{BO}^{\mathsf{S}}_i &= \left[P_t L_{\mathsf{S}i}|h_{\mathsf{S}i}|^2  + z \geq \gamma_{\mathsf{CS}} \right]\nonumber\\
\mathsf{BO}^{\mathsf{S}\mathsf{R}}_i &=\left[P_t L_{\mathsf{S}i}|h_{\mathsf{S}i}|^2+P_t L_{\mathsf{R}i}|h_{\mathsf{R}i}|^2  + \sum_{k\in\mathcal{I}_2} P_t L_{ki}|h_{ki}|^2 + z \geq \gamma_{\mathsf{CS}}\right] \nonumber
\end{align}
\noindent where $[\cdot]$ represents the Iverson bracket. Additionally, $P_t$ is a constant representing transmission power, $L_{ij}$ is a multiplicative factor that reduces power according to path loss between nodes $i$ and $j$, $z$ is a constant representing the thermal noise power in each radio, $\gamma_\mathsf{DET}$ represents a threshold for the minimum $\mathsf{SINR}$ required to decode a reception, and $\gamma_{\mathsf{CS}}$ represents a power threshold for carrier-sensing. Finally, the $\mathcal{I}$ subsets represent other potential transmitters in the network (including other relays) as defined by
\begin{align}
\mathcal{I}_1 &= \mathcal{N}\setminus\{\mathsf{S},\mathsf{D},\mathsf{R},i,j\}\nonumber\\
\mathcal{I}_2 &= \mathcal{N}\setminus\{\mathsf{S},\mathsf{D},\mathsf{R},i\}\nonumber.
\end{align}
The $\mathsf{Cons}(0)$ protocol can simply be stated as $X^{\mathsf{Cons}(0)}_\mathsf{R} = \mathsf{Rx}$ since the relay never transmits.

These protocols ensure that the relay is disabled whenever it would cause a deferral or collision event in surrounding flows. As such, they guarantee zero reduction in spatial reuse. 

\noindent {\bf Greedy Protocols:}

Similarly, the $\mathsf{Greed}(2)$ and $\mathsf{Greed}(0)$ policies can be directly translated into protocols that operate on instantaneous $\mathsf{SINR}$ measurements.

\begin{algorithm}[h!]
\label{al:greed2}
$\mathcal{N} = \{0,1,2,\ldots,N-1\}$\\
$X^{\mathsf{Greed}(2)}_\mathsf{R} = \mathsf{Tx}$\\
\For{$i \in \mathcal{N}\setminus \{\mathsf{S},\mathsf{D},\mathsf{R}\}$}{
\For{$j \in \mathcal{N}\setminus \{i,\mathsf{S},\mathsf{D},\mathsf{R}\}$}{
\If{($\overline{\mathsf{BO}^{\mathsf{S}}_i}$ and $\mathsf{BO}^{\mathsf{S}\mathsf{R}}_i$ and $\overline{\mathsf{COL}^{\mathsf{S}}_{ij}}$ and $\mathsf{COL}^{\mathsf{S}\mathsf{R}}_{ij}$) or $\mathsf{COL}_{\mathsf{S}\mathsf{R}\mathsf{D}}$}{$X^{\mathsf{Greed}(2)}_\mathsf{R} = \mathsf{Rx}$}
}
}
\caption{$\mathsf{Greed}(2)$}
\end{algorithm}

Protocol~\ref{al:greed2} formally specifies the $\mathsf{Greed}(2)$ behavior. The collision and backoff terms are
\begin{align}
\mathsf{COL}_{\mathsf{S}\mathsf{R}\mathsf{D}} &=\left[\frac{P_t L_{\mathsf{S}\mathsf{D}} |h_{\mathsf{S}\mathsf{D}}|^2 + P_t L_{\mathsf{R}\mathsf{D}} |h_{\mathsf{R}\mathsf{D}}|^2}{ z } < \gamma_\mathsf{DET}\right] \nonumber\\
\mathsf{COL}^{\mathsf{S}}_{ij} &=\left[\frac{P_t L_{ij} |h_{ij}|^2}{P_t L_{\mathsf{S}j}|h_{\mathsf{S}j}|^2  + z } < \gamma_\mathsf{DET}\right] \nonumber\\
\mathsf{COL}^{\mathsf{S}\mathsf{R}}_{ij} &=\left[\frac{P_t L_{ij} |h_{ij}|^2}{P_t L_{\mathsf{S}j}|h_{\mathsf{S}j}|^2 + P_t L_{\mathsf{R}j}|h_{\mathsf{R}j}|^2 + z } < \gamma_\mathsf{DET}\right] \nonumber\\
\mathsf{BO}^{\mathsf{S}}_i &= \left[P_t L_{\mathsf{S}i}|h_{\mathsf{S}i}|^2  + z \geq \gamma_{\mathsf{CS}} \right]\nonumber\\
\mathsf{BO}^{\mathsf{S},\mathsf{R}}_i &=\left[P_t L_{\mathsf{S}i}|h_{\mathsf{S}i}|^2+P_t L_{\mathsf{R}i}|h_{\mathsf{R}i}|^2  +  z \geq \gamma_{\mathsf{CS}}\right] \nonumber
\end{align}\noindent where all components share the same definitions.
The $\mathsf{Greed}(0)$ protocol can simply be stated as $X^{\mathsf{Greed}(0)}_\mathsf{R} = \mathsf{Tx}$ since it makes no effort to defer any of its transmissions.

The greedy protocols will increase the rate of the cooperative flow but will do so at some cost to surrounding flows.

\subsection{Custom Simulator}

To evaluate the distributed protocols, we construct a custom network simulator based on the well-known ns-2 simulation environment. The 802.11 extension to ns-2 contributed a number of enhancements for wireless applications~\cite{chen2007overhaul}. We have added significant extensions to include a cooperative PHY module and a path-symmetric Rayleigh fading channel module. Using our real-time FPGA implementations of cooperative protocols~\cite{Hunter:2010,warpOFDM_assilomar}, we have verified the accuracy of this simulator with actual over-the-air and channel emulator measurements.

\subsection{Performance Evaluation}
In Table~\ref{tbl:simParams}, we specify the key simulation parameters.
\begin{table}[h]
  \caption{Simulation Parameters}
\begin{center}
\begin{tabular}{ |c | c | }
\hline			
{\bf Header Rate } & BPSK (1/2 rate code)\\ \hline
{\bf Payload Rate } & 16-QAM (3/4 rate code)\\ \hline
{\bf Path loss Exponent } & 3\\ \hline
{\bf Fading } & Correlated Rayleigh\\ \hline
{\bf Doppler Freq. ($f_d)$ } & 15Hz\\ \hline  
{\bf RTS/CTS } & Disabled\\ \hline
{\bf Traffic } & CBR\\ \hline
{\bf Packet Size } & 1470 bytes\\ \hline
\end{tabular}
\label{tbl:simParams}
\end{center}
\end{table}
All other parameters in the experiment including $\mathsf{SINR}$ thresholds for packet decoding are identical to the default values specified in~\cite{chen2007overhaul}.

In Section~\ref{sct:policy}, we evaluated the various policies with a single parameter $p$ that affects the likelihood of the cooperative flow being connected to the other flow in the network. In this section, the analogous parameter is the distance between the relayed and non-relayed flow as shown in the simulation topology in Figure~\ref{fig:simSetup}.

\begin{figure}[h!]
	\centering
	\includegraphics[width=2in]{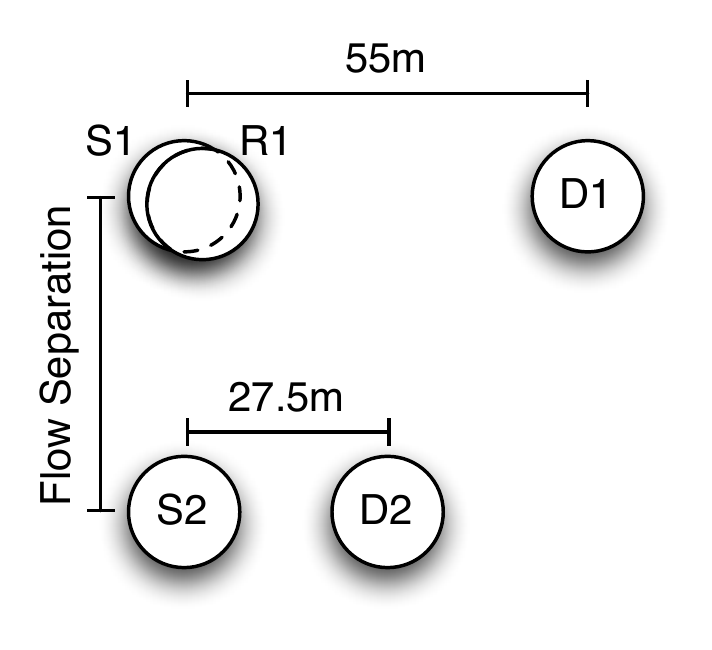}
	\caption{We vary the flow separation distance as the independent variable for the simulation.}
	\label{fig:simSetup}
\end{figure}

The cooperative flow has a large source-destination distance in order to place that flow in a fading-dominated regime (i.e. a significant number of the packet losses suffered by the destination are due to inadequate channel capacity between $\mathsf{S1}$ and $\mathsf{D1}$). The non-cooperative flow is in an interference-dominated regime where very few of its transmissions are lost due to fades. This topological selection emphasizes the negative impact of the relay on the non-cooperative flow and allows clear differences between the distributed cooperative protocols to be seen.

A useful metric for evaluating the performance of a protocol is to compare the throughput of each flow when a relay is present in the network with the throughput of each flow when there is no relay. We consider this change in throughput when a relay is present in the network.

Figure~\ref{fig:sim} shows the measured throughput difference when a relay is present and when it is not for each previously described protocol. Figure~\ref{fig:sim}\subref{fig:simFlow1} focuses on the impact of cooperation on the cooperative flow. Of note, the $\mathsf{Cons}(0)$ protocol provides no improvement over the case where the relay is absent from the network because the $\mathsf{Cons}(0)$ protocol never uses it. All other protocols, however, provide throughput improvement. In particular, the $\mathsf{Greed}(0)$ protocol (that always uses the relay) provides the most improvement at all flow separation distances. This fact is predicted by the binary network model and was shown in Figure~\ref{fig:policyPerformance}\subref{fig:help_all}. 

\begin{figure}[h!]%
\begin{center}
    \captionsetup[figure]{margin=10pt}
    \subfloat[Cooperative Flow Performance.\label{fig:simFlow1}]
    {\includegraphics[width=3in]{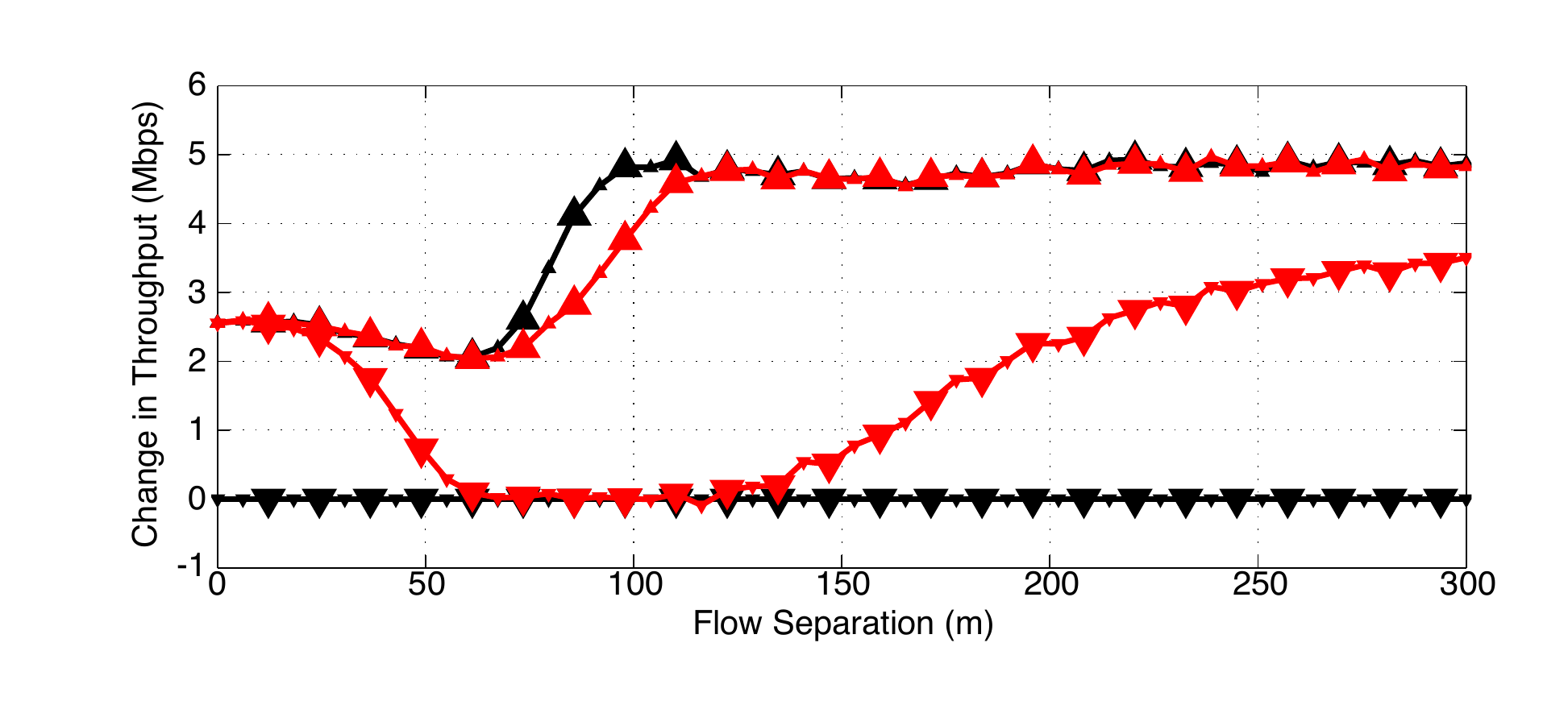}}
     \qquad
    \subfloat[Non-cooperative Flow Performance.\label{fig:simFlow2}]
    {\includegraphics[width=3in]{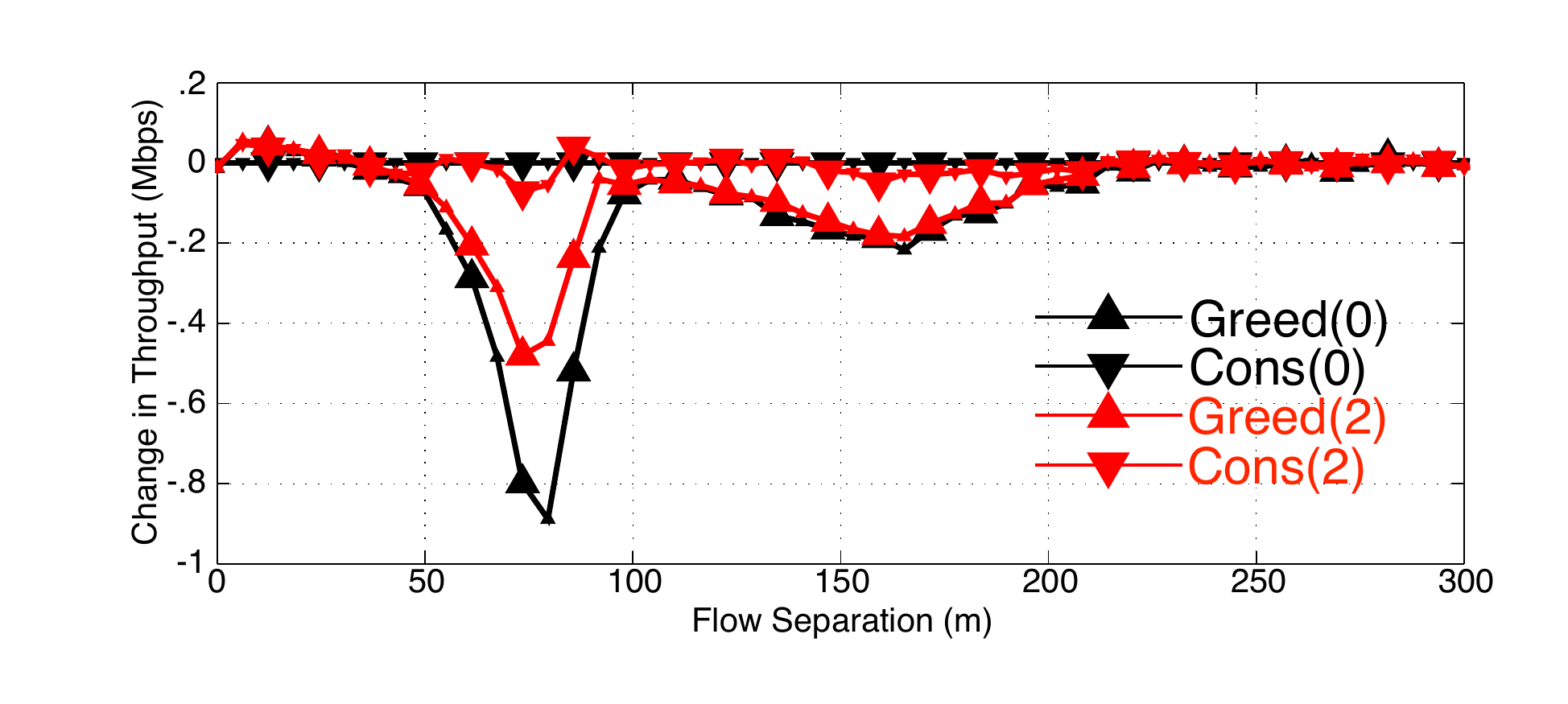}}
            \caption{The relay can assist the flow with which it is paired and harm the flow with which it is not. These effects can be balanced with protocol selection.}%
    \label{fig:sim}%
\end{center}
\end{figure}

Figure~\ref{fig:sim}\subref{fig:simFlow2} shows the impact of relaying on the non-cooperative flow. Since the relay can only increase the footprint of the cooperative flow, this means that spatial reuse can only degrade and not improve the performance of the non-cooperative flow. Again, the $\mathsf{Cons}(0)$ protocol never uses the relay so it never degrades the throughput of the non-cooperative flow. $\mathsf{Greed}(0)$, however, has two distinct regions where the harm on non-cooperative flow reaches local maxima. The reason there are two regions is that the locations where collisions and backoff deferrals each create the most impact are not necessarily the same; they depend on the many parameters specific to the simulation. Regardless, the $\mathsf{Cons}(2)$ protocol avoids {\it any} harm just as the corresponding policy predicts in Section~\ref{sct:policy}. In Figure~\ref{fig:policyPerformance}, each policy was analyzed as a function of a proxy for flow separation. Noting the similarities with the actual flow separation comparisons in Figure~\ref{fig:sim}, this confirms the binary model as a robust mechanism for the procedural generation of cooperative protocols.

\section{Conclusions}
\label{sct:conclusion}

Physical layer cooperation shows tremendous potential for performance improvement in wireless links that are able to use cooperative relays. However, for links that are unable to use relays, cooperation is a threat to their own performance due to the loss of spatial reuse caused by additional transmitters in the shared wireless medium. In this work, we have presented a policy design methodology that allows the systematic study of relay behavior for arbitrary amounts of network knowledge at the relay. Through extensive network simulations, we demonstrate the successful application of this method to distributed protocols that operate in fading environments.


\newpage

\appendices
\section{State Classification in Binary Model}
\label{sct:appendix}

Recalling that $H^B_{ij}$ and $X_i$ are binary valued, let 
\begin{align*}
\psi^B &= H^B_\mathsf{R1D2}\cdot2^0 + H^B_\mathsf{R1D1}\cdot2^1 + H^B_\mathsf{R1S2}\cdot2^2 + H^B_\mathsf{S2D1}\cdot2^3 + \\
&~~~~ H^B_\mathsf{S2D2}\cdot2^4 + H^B_\mathsf{S1S2}\cdot2^5 + H^B_\mathsf{S1D2}\cdot2^6 + X_\mathsf{S2}\cdot2^7.
\end{align*}

Each network state is labelled in the following table.

\begin{table}[h]
\scriptsize
\begin{center} 
\begin{tabular}{|c c||c c||c c||c c||c c||c c||c c||c c|}
\hline
$\psi^B$&Label&$\psi^B$&Label&$\psi^B$&Label&$\psi^B$&Label&$\psi^B$&Label&$\psi^B$&Label&$\psi^B$&Label&$\psi^B$&Label
\\
\hline \hline
0&$\mathcal{D}$&32&$\mathcal{D}$&64&$\mathcal{D}$&96&$\mathcal{D}$&128&$\mathcal{D}$&160&$\mathcal{D}$&192&$\mathcal{D}$&224&$\mathcal{D}$
\\
\hline
1&$\mathcal{D}$&33&$\mathcal{D}$&65&$\mathcal{D}$&97&$\mathcal{D}$&129&$\mathcal{D}$&161&$\mathcal{D}$&193&$\mathcal{D}$&225&$\mathcal{D}$
\\
\hline
2&$\mathcal{A}$&34&$\mathcal{A}$&66&$\mathcal{A}$&98&$\mathcal{A}$&130&$\mathcal{A}$&162&$\mathcal{A}$&194&$\mathcal{A}$&226&$\mathcal{A}$
\\
\hline
3&$\mathcal{A}$&35&$\mathcal{A}$&67&$\mathcal{A}$&99&$\mathcal{A}$&131&$\mathcal{A}$&163&$\mathcal{A}$&195&$\mathcal{A}$&227&$\mathcal{A}$
\\
\hline
4&$\mathcal{B}$&36&$\mathcal{D}$&68&$\mathcal{B}$&100&$\mathcal{D}$&132&$\mathcal{D}$&164&$\mathcal{D}$&196&$\mathcal{D}$&228&$\mathcal{D}$
\\
\hline
5&$\mathcal{B}$&37&$\mathcal{D}$&69&$\mathcal{B}$&101&$\mathcal{D}$&133&$\mathcal{D}$&165&$\mathcal{D}$&197&$\mathcal{D}$&229&$\mathcal{D}$
\\
\hline
6&$\mathcal{A} \cap \mathcal{B}$&38&$\mathcal{A}$&70&$\mathcal{A} \cap \mathcal{B}$&102&$\mathcal{A}$&134&$\mathcal{A}$&166&$\mathcal{A}$&198&$\mathcal{A}$&230&$\mathcal{A}$
\\
\hline
7&$\mathcal{A} \cap \mathcal{B}$&39&$\mathcal{A}$&71&$\mathcal{A} \cap \mathcal{B}$&103&$\mathcal{A}$&135&$\mathcal{A}$&167&$\mathcal{A}$&199&$\mathcal{A}$&231&$\mathcal{A}$
\\
\hline
8&$\mathcal{D}$&40&$\mathcal{D}$&72&$\mathcal{D}$&104&$\mathcal{D}$&136&$\mathcal{D}$&168&$\mathcal{D}$&200&$\mathcal{D}$&232&$\mathcal{D}$
\\
\hline
9&$\mathcal{D}$&41&$\mathcal{D}$&73&$\mathcal{D}$&105&$\mathcal{D}$&137&$\mathcal{D}$&169&$\mathcal{D}$&201&$\mathcal{D}$&233&$\mathcal{D}$
\\
\hline
10&$\mathcal{A}$&42&$\mathcal{A}$&74&$\mathcal{A}$&106&$\mathcal{A}$&138&$\mathcal{D}$&170&$\mathcal{D}$&202&$\mathcal{D}$&234&$\mathcal{D}$
\\
\hline
11&$\mathcal{A}$&43&$\mathcal{A}$&75&$\mathcal{A}$&107&$\mathcal{A}$&139&$\mathcal{D}$&171&$\mathcal{D}$&203&$\mathcal{D}$&235&$\mathcal{D}$
\\
\hline
12&$\mathcal{B}$&44&$\mathcal{D}$&76&$\mathcal{B}$&108&$\mathcal{D}$&140&$\mathcal{D}$&172&$\mathcal{D}$&204&$\mathcal{D}$&236&$\mathcal{D}$
\\
\hline
13&$\mathcal{B}$&45&$\mathcal{D}$&77&$\mathcal{B}$&109&$\mathcal{D}$&141&$\mathcal{D}$&173&$\mathcal{D}$&205&$\mathcal{D}$&237&$\mathcal{D}$
\\
\hline
14&$\mathcal{A} \cap \mathcal{B}$&46&$\mathcal{A}$&78&$\mathcal{A} \cap \mathcal{B}$&110&$\mathcal{A}$&142&$\mathcal{D}$&174&$\mathcal{D}$&206&$\mathcal{D}$&238&$\mathcal{D}$
\\
\hline
15&$\mathcal{A} \cap \mathcal{B}$&47&$\mathcal{A}$&79&$\mathcal{A} \cap \mathcal{B}$&111&$\mathcal{A}$&143&$\mathcal{D}$&175&$\mathcal{D}$&207&$\mathcal{D}$&239&$\mathcal{D}$
\\
\hline
16&$\mathcal{D}$&48&$\mathcal{D}$&80&$\mathcal{D}$&112&$\mathcal{D}$&144&$\mathcal{D}$&176&$\mathcal{D}$&208&$\mathcal{D}$&240&$\mathcal{D}$
\\
\hline
17&$\mathcal{D}$&49&$\mathcal{D}$&81&$\mathcal{D}$&113&$\mathcal{D}$&145&$\mathcal{C}$&177&$\mathcal{C}$&209&$\mathcal{D}$&241&$\mathcal{D}$
\\
\hline
18&$\mathcal{A}$&50&$\mathcal{A}$&82&$\mathcal{A}$&114&$\mathcal{A}$&146&$\mathcal{A}$&178&$\mathcal{A}$&210&$\mathcal{A}$&242&$\mathcal{A}$
\\
\hline
19&$\mathcal{A}$&51&$\mathcal{A}$&83&$\mathcal{A}$&115&$\mathcal{A}$&147&$\mathcal{A} \cap \mathcal{C}$&179&$\mathcal{A} \cap \mathcal{C}$&211&$\mathcal{A}$&243&$\mathcal{A}$
\\
\hline
20&$\mathcal{B}$&52&$\mathcal{D}$&84&$\mathcal{B}$&116&$\mathcal{D}$&148&$\mathcal{D}$&180&$\mathcal{D}$&212&$\mathcal{D}$&244&$\mathcal{D}$
\\
\hline
21&$\mathcal{B}$&53&$\mathcal{D}$&85&$\mathcal{B}$&117&$\mathcal{D}$&149&$\mathcal{C}$&181&$\mathcal{C}$&213&$\mathcal{D}$&245&$\mathcal{D}$
\\
\hline
22&$\mathcal{A} \cap \mathcal{B}$&54&$\mathcal{A}$&86&$\mathcal{A} \cap \mathcal{B}$&118&$\mathcal{A}$&150&$\mathcal{A}$&182&$\mathcal{A}$&214&$\mathcal{A}$&246&$\mathcal{A}$
\\
\hline
23&$\mathcal{A} \cap \mathcal{B}$&55&$\mathcal{A}$&87&$\mathcal{A} \cap \mathcal{B}$&119&$\mathcal{A}$&151&$\mathcal{A} \cap \mathcal{C}$&183&$\mathcal{A} \cap \mathcal{C}$&215&$\mathcal{A}$&247&$\mathcal{A}$
\\
\hline
24&$\mathcal{D}$&56&$\mathcal{D}$&88&$\mathcal{D}$&120&$\mathcal{D}$&152&$\mathcal{D}$&184&$\mathcal{D}$&216&$\mathcal{D}$&248&$\mathcal{D}$
\\
\hline
25&$\mathcal{D}$&57&$\mathcal{D}$&89&$\mathcal{D}$&121&$\mathcal{D}$&153&$\mathcal{C}$&185&$\mathcal{C}$&217&$\mathcal{D}$&249&$\mathcal{D}$
\\
\hline
26&$\mathcal{A}$&58&$\mathcal{A}$&90&$\mathcal{A}$&122&$\mathcal{A}$&154&$\mathcal{D}$&186&$\mathcal{D}$&218&$\mathcal{D}$&250&$\mathcal{D}$
\\
\hline
27&$\mathcal{A}$&59&$\mathcal{A}$&91&$\mathcal{A}$&123&$\mathcal{A}$&155&$\mathcal{C}$&187&$\mathcal{C}$&219&$\mathcal{D}$&251&$\mathcal{D}$
\\
\hline
28&$\mathcal{B}$&60&$\mathcal{D}$&92&$\mathcal{B}$&124&$\mathcal{D}$&156&$\mathcal{D}$&188&$\mathcal{D}$&220&$\mathcal{D}$&252&$\mathcal{D}$
\\
\hline
29&$\mathcal{B}$&61&$\mathcal{D}$&93&$\mathcal{B}$&125&$\mathcal{D}$&157&$\mathcal{C}$&189&$\mathcal{C}$&221&$\mathcal{D}$&253&$\mathcal{D}$
\\
\hline
30&$\mathcal{A} \cap \mathcal{B}$&62&$\mathcal{A}$&94&$\mathcal{A} \cap \mathcal{B}$&126&$\mathcal{A}$&158&$\mathcal{D}$&190&$\mathcal{D}$&222&$\mathcal{D}$&254&$\mathcal{D}$
\\
\hline
31&$\mathcal{A} \cap \mathcal{B}$&63&$\mathcal{A}$&95&$\mathcal{A} \cap \mathcal{B}$&127&$\mathcal{A}$&159&$\mathcal{C}$&191&$\mathcal{C}$&223&$\mathcal{D}$&255&$\mathcal{D}$
\\
\hline
\end{tabular}
\end{center}
\end{table}

\bibliographystyle{IEEEtran}
\bibliography{citations}

\end{document}